\documentclass[12pt]{iopart}

\usepackage{iopams}
\usepackage{graphicx}
\usepackage{cite}
\usepackage{color}
\usepackage{array}

\begin{document}
\title{Terahertz pulse generation by two-color laser fields with circular polarization}

\author{C. Tailliez$^{1,2}$, A. Stathopulos$^{1,2}$, S. Skupin$^3$, D. Buo\v{z}ius$^4$, I. Babushkin$^{5,6}$, V. Vai\v{c}aitis$^4$,  L. Berg{\'e}$^{1,2}$}

\address{$^1$ CEA, DAM, DIF, 91297 Arpajon, France}
\address{$^2$ Universit\'e Paris-Saclay, CEA, LMCE, 91680 Bruy\`eres-le-Ch\^atel, France}
\address{$^3$ Institut Lumi{\`e}re Mati{\`e}re, UMR 5306 Universit{\'e} Lyon 1 - CNRS, Universit{\'e} de Lyon, 69622 Villeurbanne, France}
\address{$^4$ Laser Research Center, Vilnius University, Saul{\.e}tekio 10, Vilnius LT-10223, Lithuania}
\address{$^5$ Institute of Quantum Optics, Leibniz University Hannover, Welfengarten 1 30167, Hannover, Germany}
\address{$^6$ Cluster of Excellence PhoenixD (Photonics, Optics, and Engineering-
Innovation Across Disciplines), 30167 Hannover, Germany}

\ead{stefan.skupin@univ-lyon1.fr}

\begin{abstract}
We study the influence of the polarization states of femtosecond two-color pulses ionizing gases on the emitted terahertz radiation. A local-current model and plane-wave evaluations justify the previously-reported impact on the THz energy yield and an (almost) linearly-polarized THz field when using circularly-polarized laser harmonics. For such pump pulses, the THz yield is independent on the relative phase between the two colors. When the pump pulses have same helicity, the increase in the THz yield is associated to longer ionization sequences and higher electron transverse momenta acquired in the driving field. Reversely, for two color pulses with opposite helicity, the dramatic loss of THz power comes from destructive interferences driven by the highly symmetric response of the photocurrents lined up on the third harmonic of the fundamental pulse. While our experiments confirm an increased THz yield for circularly polarized pumps of same helicity, surprisingly, the emitted THz radiation is not linearly-polarized. This effect is explained by means of comprehensive 3D numerical simulations highlighting the role of the spatial alignment and non-collinear propagation of the two colors.
\end{abstract}

\pacs{42.65.Re, 32.80.Fb, 52.50.Jm}
\submitto{\NJP}

\maketitle

\section{Introduction}
\label{sec1}

Terahertz (THz) waves are rapidly attracting interest due to their wide application range, both in imaging techniques and coherent spectroscopy \cite{Tonouchi:np:1:97,Wallace:fd:126:255,Jepsen:lpr:5:124,Berge:epl:2019} applied to various areas such as medical diagnostics, remote detection or cultural heritage to mention a few. In addition, energetic $(> 100\,\mu$J) THz pulses with GV/m field strength open promising perspectives in the development of compact THz electron accelerators \cite{Curcio:sr:8:1052} and THz-triggered chemistry \cite{Larue:prl:115:036103}, for which laser-plasma particle accelerators may supply suitable emitters in the future \cite{Hamster:prl:71:2725,Leemans:prl:91:074802,Dechard:prl:120:144801}. Imaging applications usually need narrow-band and up to $\mu$J energy radiation produced by, e.g., photoconductive antennas \cite{Suen:apl:96:141103,Ropagnol:oe:24:11299} or quantum cascade lasers \cite{Bachmann:opt:3:1087}. THz-driven electron acceleration requires in turn narrowband, but more energetic (sub-mJ) emitters, currently provided by optical rectification in crystals \cite{Vicario:prl:112:213901,Fulop:oe:22:20155}. These solid-based technologies are, however, hampered by their inherent damage threshold. By contrast, gas-plasmas excited by intense, two-color femtosecond pulses at intensities close to ionization threshold $\sim 10^{14}$ W/cm$^2$ are known to already produce significant conversion efficiency $> 10^{-4}$, relatively high field strength $\sim 0.1$ GV/m and $\sim 50-100$ THz broadband THz emissions \cite{Cook:ol:25:1210,Kress:ol:29:1120,Oh:njp:15:075002}. Generally created by the combination of a fundamental harmonic (FH) and second harmonic (SH), the generation mechanism is here the creation of photocurrents induced by the free electrons tunnel-ionized from air molecules and being accelerated by an asymmetric two-color field \cite{Kim:np:2:605,Kim:oe:15:4577,Berge:prl:110:073901}. Excited by collinearly-polarized (LP-P) pulse components, the measured THz power is highly sensitive to the relative phase difference, $\phi$, between the two colors and exhibits maxima when the value of $\phi$ is close to $\pi/2$ (modulo $\pi$) \cite{Kim:np:2:605,Li:apl:101:161104}. 

Several means for increasing the THz energy yield emitted by photocurrents have been explored. Among the laser-medium parameters exploited for this purpose, increasing the FH wavelength led to a significant growth in the THz pulse energy proportional to $\lambda_0^{4-5}$ using OPCPA lasers operating between 0.8 and 2 $\mu$m \cite{Clerici:prl:110:253901}. Comprehensive numerical simulations of this experiment demonstrated the crucial role of the relative phase initiated along the SH generation stage through a doubling crystal (e.g., BBO) and achieved in the plasma zone \cite{Nguyen:ol:44:1488}. More recently, the advent of sub-mJ, ultrafast mid-IR $(3.9\,\mu$m) and CO$_2$ $(10.6\,\mu$m) lasers inspired both numerical \cite{Fedorov:oe:26:31150,Nguyen:pra:97:063839} and experimental \cite{Jang:optica:6:1338,Koulouklidis:nc:11:292} investigations confirming the increase in the conversion efficiency over $1-2\;\%$ and $\sim0.1$ mJ energies for pump energies $< 10$ mJ. Alternative methods for increasing the THz yield may also rely on an optimum tuning of the intensity level versus the ionized gas (e.g., Ar, He) and their successive electron shells \cite{Debayle:pra:91:041801}, playing on the pump pulse duration or reducing the plasma dimensions \cite{Thiele:pre:94:063202,Buccheri:opt:2:366,Thiele:pra:96:053814}, increasing the number of colors \cite{Gonzalez:prl:114:183901,Zhou:ieeep:1:2018,Vaicaitis:jap:125:173103} or even modifying the frequency ratio between the two colors \cite{Zhang:prl:119:235001,Wang:pra:96:023844}.  Combining these techniques can then be expected to augment the THz pulse energy by at least one order of magnitude.

Despite the proven effectiveness of the previous methods, there exists a direct means to improve the THz performances in air photonics, which consists in simply changing the polarization state of the FH and SH components. Pioneering a coherent control of THz wave generation through the polarization state of the two colors, Dai et al. \cite{Dai:prl:103:023001} experimentally reported a continuous rotation of the resulting THz field polarization, always remaining linear, and the invariance of the radiation energy yield by changing the two-color relative phase $(\phi)$ when using circularly-polarized pulses. The dependency of the THz pulse energy with respect to the polarization of two-color laser pulses was later thoroughly investigated by Meng et al. \cite{Meng:apl:109:131105}. Inspired by the remarkable enhancement of the electron energy driven by tunnel ionization \cite{Yuan:pra:84:013426}, these authors experimentally evidenced from helium gas jets that circularly-polarized (CP) two-color pulses with same helicity (CP-S) could deliver $\sim 5$ times higher THz powers than their linearly-polarized, parallel (LP-P) counterparts. Despite ionization yields being reduced for CP pulses due to a factor $1/\sqrt{2}$ in the electric field amplitude, the increase in the THz power was attributed to an ionization process saturated (for completely ionized atoms) closer to the peak of the laser electric field and associated to the highest values of the acquired electron drift velocity. By contrast, CP pulses with counter-helicity (CP-C) in their SH delivered a much weaker THz yield, remaining, in the experiments \cite{Meng:apl:109:131105}, of the order of that supplied by LP pulses with orthogonally-polarized colors (LP-O). Among the four investigated polarization states (CP-S, CP-C, LP-P, LP-O), only pulses being linearly polarized underwent a strong dependency on the relative phase $\phi$. Meanwhile, the effects of the pump polarization states were also investigated numerically. In \cite{Fedorov:ppcf:59:014025} the influence of circularly-polarized two-color pulses driven either by four-wave mixing (FWM) or by photocurrents confirmed the previous dependencies of the resulting THz field on the two-color phase offset. In Ref. \cite{Song:apl:103:261102}, the generation and control of elliptically-polarized THz waves was demonstrated from air plasmas driven by few-cycle CP pulses. For in-line laser focusing, the THz polarization state was experimentally found to evolve from linear to elliptical by increasing the plasma length \cite{You:ol:38:1034}. More recently, theoretical evaluations based on an extended 3D local-current (LC) model \cite{Yousef:jmo:64:300} explored the role of the laser parameters (pulse duration, relative phase, tilt angle of linear polarization) on the radiation characteristics. O. Kosareva and co-workers \cite{Kosareva:ol:43:90,Esaulkov:fo:8:73}, by analyzing the polarization properties of the pulse harmonics and broadband THz generation from atmospheric plasmas, reported the decrease by about one order of magnitude of the THz yield provided by LP-O pulses compared to LP-P's. In \cite{Kosareva:ol:43:90}, abrupt changes in the THz polarization were observed from an angle $\sim 85^{\circ}$ between the FH and SH polarization axis and a weak ellipticity was sufficient to drive efficiently an elliptical THz radiation.

From the numerical point of view, solving the quantum time-dependent Schr{\"o}dinger equation (TDSE) based on the single-atom response \cite{Dai:prl:103:023001,Meng:apl:109:131105,Tulsky:pra:98:053415} could figure out partial features on the previous properties. However, we are still missing a thorough explanation of the changes induced into the characteristic laser-driven ionization steps and the electron transverse momenta by two-color pumps with different polarization states, and of their impact on the THz conversion efficiency. Such modifications can be cleared up by means of the local-current (LC) model and the microscopic description of the THz spectra built by photocurrents in the tunnel-ionization regime \cite{Babushkin:njp:13:123029}, which is one of the goals of the present work. Here, we revisit the effects of two-color laser fields with elliptical polarization on the THz yield and on the polarization state of the THz radiation. The LP-O configuration has already been justified by a vanishing of the (dominant) FH component in the electron density spectrum and a THz radiation only conveyed by the (minor) SH component \cite{Thiele:pra:96:053814,Esaulkov:fo:8:73,Nguyen:njp:20:033026}. Therefore, emphasis will be mainly given to CP-S, CP-C and LP-P pump configurations throughout the present analysis. We shall also examine the stability of both the THz energy yield and the radiation polarization state experimentally and by comprehensive numerical simulations based on a unidirectional solver \cite{Kolesik:pre:70:036604}. In this work, we confirm and explain the high THz energy level reached when using CP-S pulses, but display evidence of the fragility of CP-C pump pulses in totally inhibiting THz emission. Special attention is further paid to the impact of the alignment of the two colors on the generated THz field and its polarization.

The paper is organized as follows. In Section \ref{sec2} we derive basic evaluations of vectorial THz components excited through FWM and photocurrents by CP and LP pulses and comment on their validity in regards to (0+1)-dimensional LC computations. This analysis explains the change in the density steps and ionization duration together with those in the electron drift velocity experienced when passing from linearly  to circularly polarized pumps. Section \ref{sec3} confirms the previous expectations from a vectorial, (3+1)-dimensional (3D) unidirectional pulse propagation model (vectorial UPPE) used to numerically simulate gas jet experiments in argon (short plasma lengths) that preserve the main predictions of the LC model. Section \ref{sec4} reports experimental data on CP-S, CP-C, LP-O and LP-P using a two-arm setup allowing an individual control of the two colors. Surprisingly, in our setup CP-S pulses produce THz radiation with no particular polarization direction. These unexpected changes are justified by the wide range of values taken by the relative phase of the two colors due to their non-collinear propagation, which produces ''young moon'' like field distributions in the THz far field.

\section{Plane wave and local-current estimates}
\label{sec2}

To understand the effect of the pump polarization states onto the THz radiated field and energy, we analyze the nonlinear source terms of the medium that can serve as efficient converters into the THz domain. For the sake of simplicity, we shall discard Raman-delayed nonlinearities induced by ro-vibrational transitions of air molecules. As already shown in \cite{Nguyen:oe:25:4720}, such nonlinearities are barely influential on the THz performances. We thus consider an instantaneous Kerr response modeling FWM through the associated nonlinear polarization vector:
\begin{equation}
\label{PKerr}
{\vec P}_{\rm Kerr}(t) =\epsilon_0 \chi^{(3)} E^2 (t) {\vec E}(t),
\end{equation}
where $\epsilon_0$ is the vacuum dielectric constant, $\chi^{(3)}$ the third-order susceptibility defined at the carrier FH frequency $\omega_0 = 2\pi c/\lambda_0$ ($c$ is the speed of light in vacuum), and $\vec{E}$ is the real-valued electric field vector. Besides optical nonlinearities, the plasma response of weakly ionized gases with neutral density $N_a$ is classically modeled by the electron source equation:
\begin{equation}
\label{Ne}
\partial_t N_e = W(E)(N_a - N_e),
\end{equation}
where $N_e$ denotes the electron density, $W(E)$ is the ionization rate only depending on the length of the electric field vector $E(t) \equiv |{\vec E}(t)|$. In the present study this rate is given by the instantaneous rate from Ammosov-Delone-Krainov (ADK) theory \cite{Ammosov:spjetp:64:1191,Thomson:lpr:1:349} reducing for hydrogenoid atoms to the well-known quasi-static tunneling (QST) rate \cite{Landau:QuantMech:1965}:
 \begin{equation}
\label{W}
W[E(t)] = \frac{\alpha}{E(t)} \mbox{e}^{- \frac{\beta}{E(t)}},
\end{equation}
where the constants $(\alpha,\beta)$ can be found defined in, e.g., \cite{Gonzalez:jpb:47:204017}.

Without loss of generality, we shall consider a vectorial Gaussian laser field: 
\begin{equation}
\label{input}
 \vec{E}_{L}(t)  =  \sum_{j=1,2}  \frac{E_{0,j}}{\sqrt{1+\rho_j^2}}
 \left(\begin{array}{c}
 \cos(j\omega_0t+\phi_j) \\
 \rho_j \cos(j\omega_0t+\phi_j + \theta_j)
\end{array}\right) \mathrm{e}^{-2 \ln 2 \frac{t^2}{\tau^2_{j}}},
\end{equation}
where the small longitudinal component is neglected. The polarization states of FH ($j=1$) and SH ($j=2$) field with amplitude $E_{0,j}$ are controlled by their respective ellipticities $\rho_j$ and phase angles $\theta_j$. The carrier envelope phase can be set by the phase offsets $\phi_j$ for each color, while the ratio $r \equiv E_2^2/E_0^2$ denotes the SH intensity fraction with $E_0=\sqrt{E_1^2+E_2^2}$ being the overall maximum laser field. The four laser configurations of particular interest are then LP-P: $\rho_j=0$, CP-S: $\rho_j=1$, $\theta_j=\pm \pi/2$, CP-C: $\rho_j=1$, $\theta_1=-\theta_2=\pm\pi/2$ and LP-O: $\rho_1=0,\,\rho_2=+\infty$, $\theta_j=0$. For the sake of conciseness, we shall restrict the phase offsets to $\phi_1 = 0,\,\phi_2 = \phi$ and assume the same FHWM duration $\tau=\tau_j$ for both colors if not stated otherwise. Unless other fractions are addressed, we shall also consider a generic SH intensity fraction of $r=10\%$ in every pump pulse configuration. 

\begin{figure}[ht]
\centering \includegraphics[width=\columnwidth]{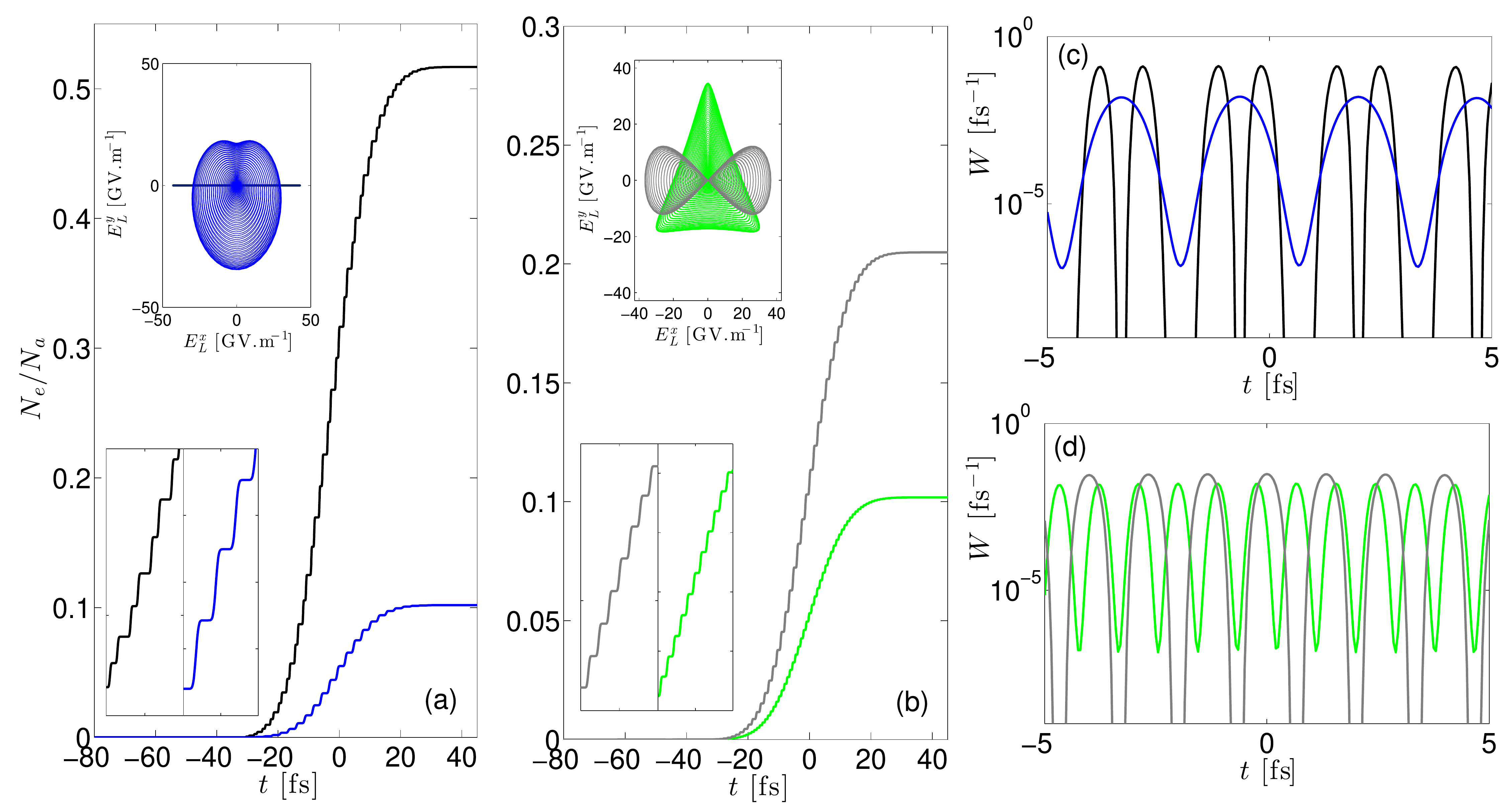}
\vskip 0.5cm
\centering \includegraphics[width=\columnwidth]{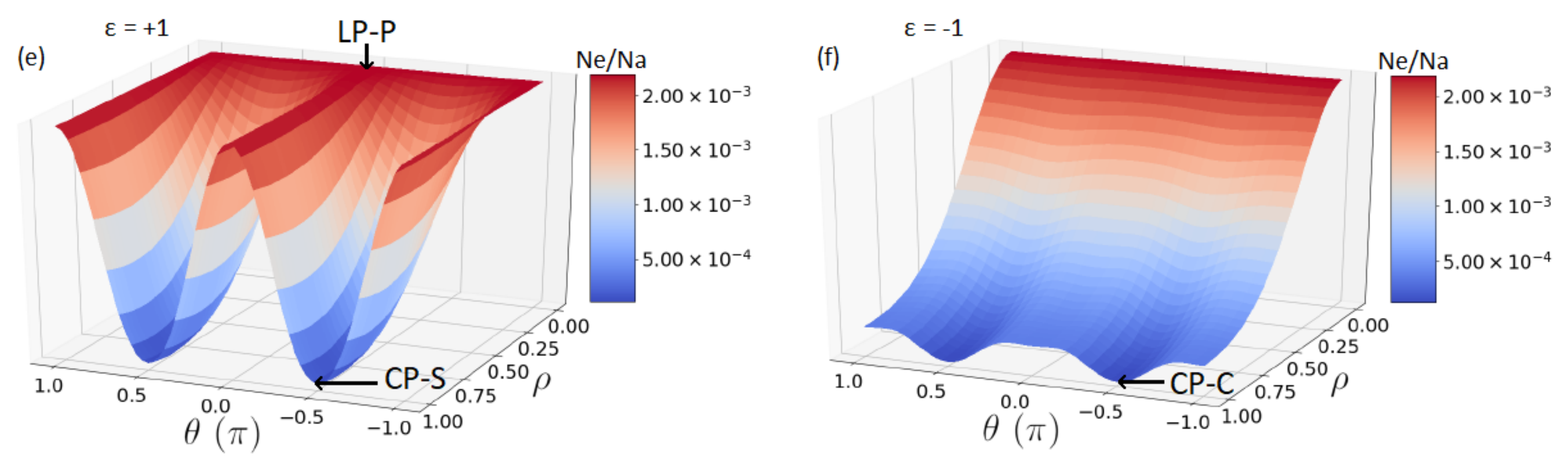} 
\caption{(a,b) Electron density $N_e(t)$ developed along the laser temporal profile in (a) the LP-P (black) and CP-S cases (blue curves), and (b) the CP-C (green) and LP-O cases (gray curves) for two-color, 60-fs Gaussian pulses with 800-nm FH and $200$ TW/cm$^2$ intensity ionizing argon at ambient pressure. Top insets detail the field polarization patterns. Bottom insets zoom in the steplike increase in $N_e(t)$ over a 10-fs time window. (c,d) Corresponding QST ionization rates. (e,f) Density maps of maximum ratios $N_e/N_a$ for finite ellipticities $\rho_1 = \rho_2 \equiv \rho \leq 1$ and $\theta_1 = \theta_2 \equiv \theta$ for two colors with (e) same helicity and (f) opposite helicity.}
\label{Fig1}
\end{figure}

Figure \ref{Fig1} displays illustrative examples of field patterns, ionization yields and QST rates for two-color 60-fs Gaussian pulses with 200 TW/cm$^2$ intensity ionizing argon at ambient pressure for the four LP-P, CP-S, CP-C and LP-O pulses with $\lambda_0 = 800$ nm and relative phase $\phi = \pi/2$ that promotes an optimally-emitting plasma zone in the LP-P case. Higher electron density is reached with a larger field strength for the linearly-polarized pulses. By contrast, the CP-S pulse develops a lower electron yield through longer and more regular ionization steps lined up on the FH frequency  [see Fig. \ref{Fig1}(a), bottom inset and Fig. \ref{Fig1}(c)]. The CP-C pulse induces a 3rd-harmonic periodic growth in the electron density connected to the $3\omega_0$ periodicity of its ionization response [see Fig. \ref{Fig1}(b), bottom inset and Fig. \ref{Fig1}(d)]. The top insets in Fig. \ref{Fig1} describe the pump field polarization patterns. Note the flattened CP-S contours due to a non-zero SH intensity ratio and the perfect triangular contour of CP-C pulses offering a very symmetric distribution at their field maxima where ionization takes place. As can be seen from Figs. \ref{Fig1}(c,d), the ionization rate for Ar never vanishes for CP pulses. Figures \ref{Fig1}(e,f) detail the density maps for the same intensity value and pulse durations in the further parameter ranges of interest, i.e., $0 \leq \rho_1 = \rho_2 \equiv \rho \leq 1$ and $\theta_1 = \theta_2 \equiv \theta$. These maps reveal important variations in the ionization yield of Ar when modifying the pump polarization states. They will enable us to better appreciate the effective gain in the conversion efficiency achieved when these polarization states are changed.

Discarding the LP-O case (already treated in \cite{Thiele:pra:96:053814,Esaulkov:fo:8:73,Nguyen:njp:20:033026}), we now focus on the LP-P, CP-S and CP-C configurations. For the following analysis the laser field (\ref{input}) can be simplified by involving a reduced number of parameters as
\begin{equation}
\label{input_2}
{\vec E}_L(t) = \frac{E_0 \mathrm{e}^{-2 \ln 2 \frac{t^2}{\tau^2}}}{\sqrt{1+\rho^2}} \left [ \sqrt{1-r} \left(\!\begin{array}{>{\scriptstyle}c}
 \cos(\omega_0t) \\
 \rho \cos(\omega_0t + \theta)
\end{array}\!\right) 
+ \sqrt{r} \left(\!\begin{array}{>{\scriptstyle}c}
 \cos(2\omega_0t + \phi) \\
 \rho \epsilon \cos(2\omega_0t + \theta + \phi)
\end{array}\!\right) \right].
\label{input2_2}
\end{equation}
The LP-P, CP-S and CP-C pump configurations are recovered by simply setting $\rho = 0$, $(\rho = 1,\,\epsilon = 1,\theta = -\pi/2)$ and $(\rho=1,\,\epsilon = -1,\theta = -\pi/2)$, respectively. Keeping, however, the parameters $\rho$, $\theta$, and $\epsilon$ covering a broader range of values will allow us to study what happens if pump polarizations are imperfect, e.g, a slight ellipticity is introduced.

Let us start our analysis by using simple plane wave arguments, i.e., assuming the limit $\tau \rightarrow +\infty$ in Eq. (\ref{input_2}). First clues on the nonlinear converters' efficiency can be obtained by just looking at the quasi-DC (low-frequency) contributions extracted from the optical or plasma nonlinearities. For CP or LP-P pulses the electric field length expands as
\begin{eqnarray}
\label{length2}
E_L(t) & = \frac{E_0}{\sqrt{2(1+\rho^2)}} \Big \{ 1 + \rho^2 + 2 \sqrt{r(1-r)} (1+\epsilon\rho^2) \cos{(\omega_0 t + \phi)} \nonumber \\
& \quad + (1-\rho^2) (1-r) \cos{(2\omega_0 t)} + 2 \sqrt{r(1-r)} (1- \epsilon\rho^2) \cos{(3\omega_0 t + \phi})   \nonumber \\
& \quad + (1-\rho^2) r \cos{(4\omega_0 t + 2 \phi)} \Big \}^{1/2},
\end{eqnarray}
which contains components oscillating at the FH, SH, third and fourth harmonic frequencies.
This expression readily leads to the DC Kerr polarization vectors:
\begin{equation}
{\vec P}_{\rm Kerr}^{\rm LP-P} = \frac{3 \epsilon_0}{4} \chi^{(3)} E_0^3 \sqrt{r}(1-r) \left( \begin{array}{c} \cos{\phi} \\ 0 \end{array} \right), \label{KLP}
\end{equation}
\begin{equation}
{\vec P}_{\rm Kerr}^{\rm CP} = \frac{\epsilon_0}{2^{5/2}} \chi^{(3)} E_0^3 \sqrt{r}(1-r) (1+\epsilon) \left( \begin{array}{c} \cos{\phi} \\ - \sin{\phi} \end{array} \right), \label{KCP}
\end{equation}
indicating that two-color CP pulses may produce a THz radiation being linearly polarized along the angle $-\phi$ when it is driven by the Kerr response only. More involved features can be expected from pulsed beams for which the Kerr-driven THz yield will be evaluated from filtering in frequency the spectrum of ${\vec E}_{\rm Kerr}^{\rm THz} \propto \partial_t^2 {\vec P}_{\rm Kerr}$ \cite{Berge:prl:110:073901,Borodin:ol:38:1906}. Note that we discarded the factors $(1/3,2/3)$ affecting $n_2$ in the cross- and self-phase modulation terms for CP pulses \cite{Agrawal:NFO:01,Berge:pd:176:181}, as those will not change the main conclusion below.  

\begin{figure}
\centering 
\includegraphics[width=\columnwidth]{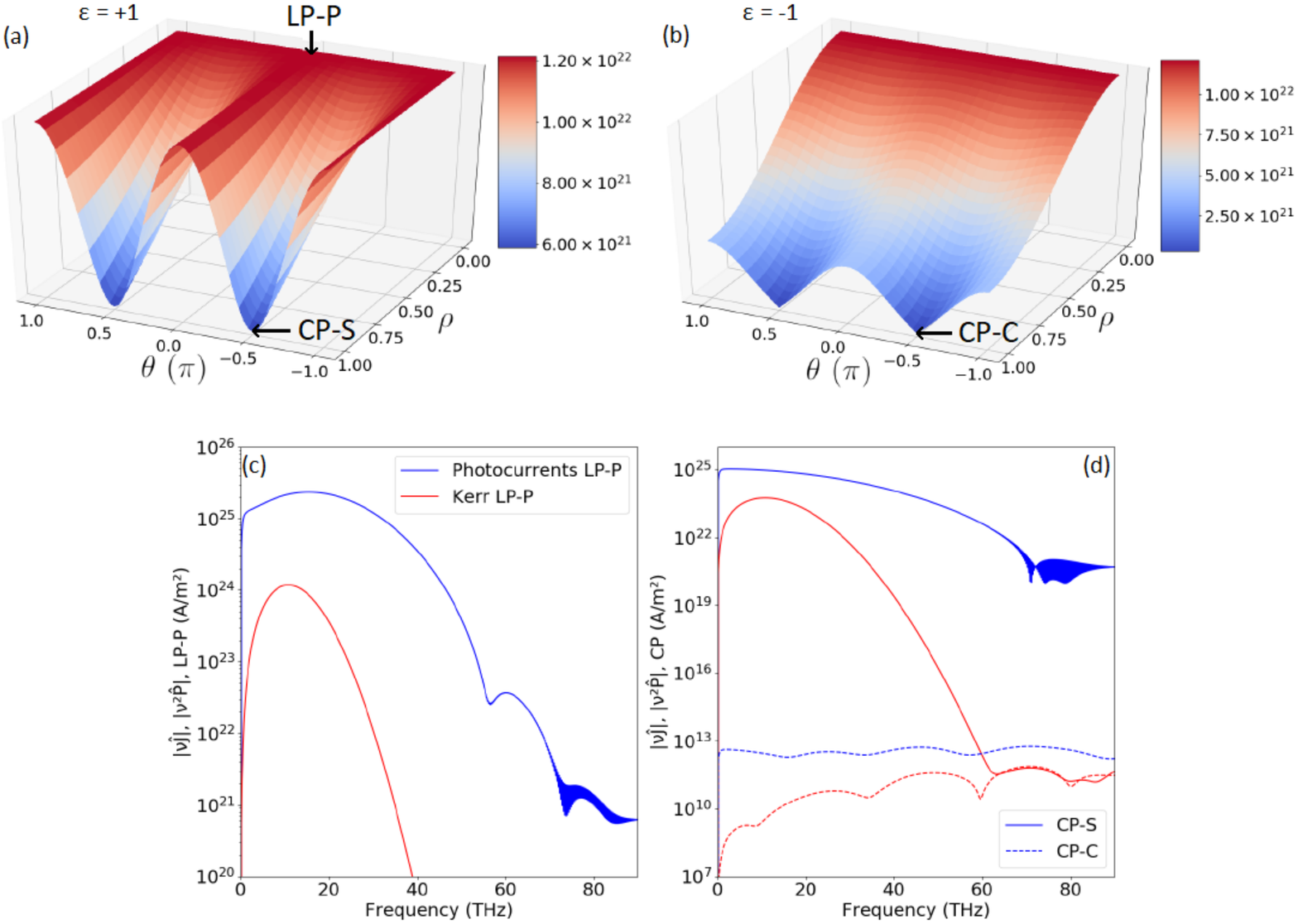}
\caption{Mapping over $(\rho,\theta)$ of max$_t \, |\partial_t^2 {\vec P_{\rm Kerr}}|$ in A/m$^2$ s$^{-1}$ units filtered out in the frequency domain $\nu \leq \nu_{\rm co} = 90$ THz for the cases (a) $\epsilon = + 1$ and (b) $\epsilon = -1$ with $\phi = 0$ in two-color, 60-fs Gaussian pulses with Kerr index of $10^{-19}$ cm$^2$/W. (c,d) THz spectra for (c) LP-P, CP-S and (d) CP-C pulses comparing the efficiency of the Kerr ($\phi = 0$) and photocurrent ($\phi = \pi/2$) nonlinearities obtained from the LC model (blue curves: photocurrents, red curves: Kerr nonlinearity). The electron-neutral collision rate is $\nu_c = 2.85$ ps$^{-1}$.}
\label{FigAlex1}
\end{figure}

On the other hand, to evaluate the efficiency of photocurrents, the THz field ${\vec E}_{\mathrm{THz}}$ is extracted according to the LC model \cite{Kim:np:2:605,Babushkin:prl:105:053903} by filtering the secondary field ${\vec E}_{J} = g \partial _t {\vec J}$ emitted from the current density $\vec{J}$ induced by free electrons, $g$ being a geometrical factor originating from Jefimenko's theory \cite{Jefimenko:EM:66}. At moderate intensities $< 10^{15}$ W/cm$^2$, the temporal shape of ${\vec J}$ is given by the cold-plasma kinetic equation \cite{Kim:np:2:605}:
\begin{equation}
\label{J}
\partial_t {\vec J} + \nu_c {\vec J} = \frac{e^2}{m_e} N_e {\vec E},
\end{equation}
where $e$ and $m_e$ are the electron charge and mass, respectively, and $\nu_c = 2.85$ ps$^{-1}$ denotes the electron-neutral collision rate associated to $\sim 350$ fs collision time. Following the method exploited in Ref. \cite{Nguyen:njp:20:033026}, the THz waveform follows from the low-frequency (quasi-DC) contribution of the product $N_e(t) {\vec E}(t)$, where ${\vec E}(t) = {\vec E}_L(t)$. The electron density driven by the vectorial laser field (\ref{input_2}) can be approximated in the limit $N_e \ll N_a$ and along the slope where $W(E)$ increases with $E$ by $N_e^L(t) \simeq N_a \int_{-\infty}^t W[E_L(t')] dt' \propto \int_{-\infty}^t E_L^2(t') dt'$. Given the electric field length (\ref{length2}), the THz field polarization state estimated from a plane-wave theory expresses in the collisionless limit $\nu_c \rightarrow 0$ as:
\begin{equation}
{\vec E}_{\rm PC}^{\rm LP-P} \propto E_0^3 \sqrt{r}(1-r) \left( \begin{array}{c} \sin{\phi} \\ 0 \\ \end{array} \right), \label{PCLP}
\end{equation}
\begin{equation}
{\vec E}_{\rm PC}^{\rm CP} \propto E_0^3 \sqrt{r}(1-r) (1+\epsilon) \left( \begin{array}{c} \sin{\phi} \\ \cos{\phi} \\ \end{array} \right). \label{PCCP}
\end{equation}
The index ''PC'' here refers to the photocurrent source. From these expressions, we justify (i) the linear polarization of the THz radiation expected when using CP pump pulses, (ii) the invariance of the THz energy $\propto \int E_{\rm CP}^2(t') dt'$ with respect to the phase offset $\phi$, and (iii) the vanishing of the THz power in the CP-C configuration $(\epsilon = -1)$.

\begin{figure}
\centering 
\includegraphics[width=\columnwidth]{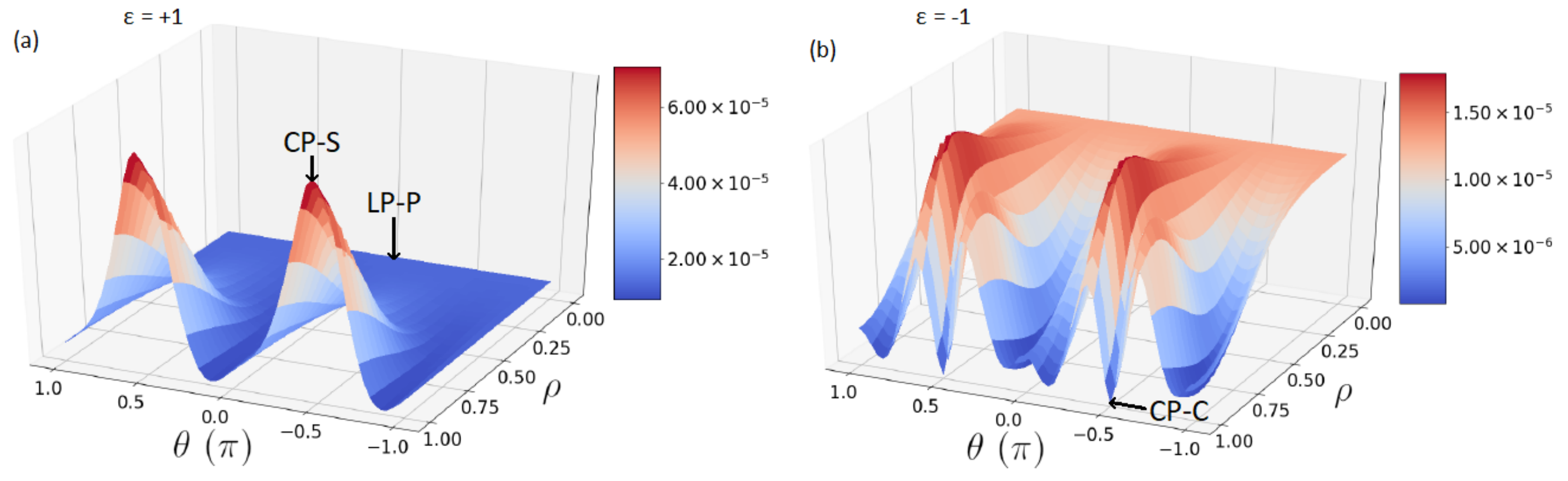} 
\caption{Mapping over $(\rho,\theta)$ of the laser-to-THz conversion efficiency as defined by Eq. (\ref{etaTHz}) from the photocurrents constrained to the same final electron density and for the same pulse and medium parameters as in Fig. \ref{Fig1}. (a) $\epsilon = + 1$, (b) $\epsilon = -1$.}
\label{FigAlex2}
\end{figure}  
For comparison, Figs. \ref{FigAlex1}(a,b) provide a mapping of the maximum-in-time of the Kerr source term filtered in the low-frequency domain $\nu \leq 90$ THz and plotted in the plane $(\rho, \theta)$ for 100 TW/cm$^2$, 60-fs two-color Gaussian pulses with $\phi = 0$. This intensity value is taken as representative of a transition between Kerr self-focusing and plasma generation in argon along a self-channeling dynamics. Several features emerge from these graphics. First, the Kerr contribution in Fig. \ref{FigAlex1} for $\epsilon = 1$ displays similar differences $\approx 2$ in the amplitude factors as Eqs. (\ref{KLP}) and (\ref{KCP}) when comparing Kerr-driven LP-P and CP-S efficiencies [Fig. \ref{FigAlex1}(a)]. Second, the THz signal is almost extinguished for a CP-C pump pulse [Fig. \ref{FigAlex1}(b)] as expected from Eq. (\ref{KCP}), the residual signal being attributed to envelope effects. For further comparison, Fig. \ref{FigAlex1}(c,d) display the THz spectra ($\nu \leq 90$ THz) of both Kerr ($\phi = 0$) and photocurrent ($\phi = \pi/2$) source terms for the same pump pulses. One can observe that the photocurrent efficiency is much higher by one to two orders of magnitude than that of the Kerr source in every configuration. This property agrees with current expectations in this intensity range \cite{Nguyen:oe:25:4720}. Since higher intensities will be attained in the coming comprehensive simulations and experiments, we shall henceforth drop the influence of the Kerr response in the following analysis. 

As highlighted by Figs. \ref{Fig1}(e) and \ref{Fig1}(f) different electron yields are attained when changing the polarization states at constant intensity. To properly apprehend the relative variations in the laser-to-THz conversion efficiency only caused by a change in the pump polarization state, it is more appropriate to remove any related increase in the electron density. Let us recall that, in practical situations, plasma defocusing in three-dimensional propagation geometries usually clamps the achieved peak intensity, which, by feedback, constrains the maximum electron density to comparable levels \cite{Nguyen:oe:25:4720}. Therefore, we henceforth opt for normalizing the THz amplitude $E_{\rm THz} \propto \partial_t J|_{\rm THz} \propto N_e^L E_L$ by the asymptotic value of $N_e(t)$ reached at large times. Because the geometrical factor $g$ introduced above is undetermined, we pick it as providing a convenient normalization with respect to the maximum electron density (see also \cite{Nguyen:njp:20:033026}):
\begin{equation}
\label{g}
g = \frac{m_e}{e^2\max_t N_e^L}.
\end{equation}
To foresee the best laser configurations, an estimate of the laser-to-THz conversion efficiency is thus defined by
\begin{equation}
\label{etaTHz}
\eta_{\rm THz} \equiv \int_{-\omega_{co}}^{\omega_{co}} |{\widehat E}_J|^2 d\omega/\int_{-\infty}^{+\infty} |{\widehat E}_L|^2 d\omega,
\end{equation}
where the $\,\widehat{}\,$ symbol denotes Fourier transform, the numerator is computed in the frequency window $\omega \leq \omega_{co}$ and $g$ is given by Eq. (\ref{g}).  

Figures \ref{FigAlex2}(a,b) show this conversion efficiency. For comparable electron density levels, maximum generation is reached for CP-S pulses in Fig. \ref{FigAlex2}(a), as expected. By contrast, CP-C pulses [see Fig. \ref{FigAlex2}(b)] supply a peculiar configuration extinguishing the THz signal only inside a narrow zone in the plane $\rho \simeq 1,\,\theta \simeq \pm \pi/2$. A small deviation from this narrow region should thus lead us to recover THz performances comparable with those reached with an LP-P or CP-S state.

To explain the gain in the THz yield when passing from LP-P to CP-S polarization states, we now perform a microscopic description of the photocurrents accounting for the pulsed nature of the laser beams. First, let us observe that, since the ionization rate $W(E)$ only depends on the length of the electric field vector, its extrema are reached at the instants: $\omega_0 t_n \approx n \pi - 2 \sqrt{r} (-1)^n \sin{\phi}/\sqrt{1-r}$ for LP-P in the limit $r \ll 1$ \cite{Babushkin:njp:13:123029}, $\omega_0 t_n = 2 n \pi - \phi$ for CP-S pulses and $\omega_0 t_n = 2n \pi/3 - \phi/3$ for CP-C ones with $n$ being an integer. It is easy to guess from Eq. (\ref{length2}) that circularly-polarized pulses with same helicity should promote a longer ionization lined up on the FH period alone, unlike the LP pulses that mix up FH and SH frequencies and trigger shorter ionization events.

Following Ref. \cite{Babushkin:njp:13:123029}, we can then sort out the vectorial high-frequency (${\vec J}_A$) and low-frequency (${\vec J}_B$) current components from Eq. (\ref{J}) as ${\vec J}(t) = {\vec J}_A(t) + {\vec J}_B(t)$ with
\begin{eqnarray}
\label{current}
{\vec J}_A(t) &=& - e \sum_n \delta N_n {\vec v}_f(t) H_n(t-t_n), \\
{\vec J}_B(t) &=& e \sum_n \delta N_n {\vec v}_f(t_n) \mbox{e}^{- \nu_c(t-t_n)} H_n(t-t_n), \label{current_2}
\end{eqnarray}
where $\delta N_n$ is the elementary ionization step formed at the $n$th ionization instant, $H_n(t-t_n) = \frac{1}{2} (1 + \mbox{erf}[(t-t_n)/\tau_n^{\rm ion}])$ models a steplike function,
\begin{equation}
\label{taun}
\tau_n^{\rm ion} = \sqrt{2} E_L(t_n)/\sqrt{\beta \partial_t^2 E|_{t=t_n}}
\end{equation}
is the ionization time scale and ${\vec v}_f(t)$ represents the drift velocity in the $(x,y)$ plane of a free electron created at $t = -\infty$:
\begin{equation}
\label{vf}
{\vec v}_f(t) = - \frac{e}{m_e} \int_{-\infty}^t {\vec E}_L(t') \mbox{e}^{-\nu_c(t-t')} dt'.
\end{equation}
Applying the Fourier analysis of Ref. \cite{Gonzalez:prl:114:183901} we extract the radiated field $E_J$ from the frequency window $\nu \leq \nu_{co} = \omega_{co}/2\pi$ with $\nu_{co} = 90$ THz, where the low-frequency radiation is mainly carried out by ${\vec E}_B^{THz} \propto \partial_t {\vec J}_B|_{\rm THz}$. This quantity is directly proportional to the kicks in the electron transverse momenta $\propto v_f(t_n)$ induced at each ionization event. Expressions of ${\vec v}_f(t_n)$ are detailed in the Appendix. It is found, in particular, that the electron velocity at $t=t_n$ with CP pulses directly depends on the dominant FH amplitude unlike the LP-P pulses fostering a THz field driven by the SH component only. As discussed in the Appendix, the SH intensity fraction $r$ must not be too small, for an electron steplike increase to make sense. Under this condition, the electron drift velocity cumulated from all ionization events is higher with CP-S pulses. This difference directly impacts the THz spectrum mainly determined by 
\begin{equation}
\label{Sp}
{\widehat {\vec E}}^J(\omega) = \frac{ge}{\sqrt{2\pi}} \sum_{n = 1}^N \delta N_n {\vec v}_f(t_n) \mbox{e}^{i \omega t_n},
\end{equation}
where $N$ denotes the total number of ionization events. Since the elementary density step barely varies with the ionization index $n$ [see Figs. \ref{Fig1}(a,b)], we assume $\delta N_n = \delta N$ for all $n$.  Comparing THz fields with equal ionization yield requires to fix the product $N\delta N$ constant for all configurations of interest. Using the above-recalled ionization instants into ${\vec v}_f(t_n)$, the THz spectrum for LP-P pulses expresses, in the limits $\omega /\omega_0 \ll 1$ and $N \gg 1$, as \cite{Babushkin:njp:13:123029}
\begin{equation}
\label{SpLP}
{\widehat E}^J_{\rm LP-P}(\omega) = \frac{3 g e^2 N \delta N E_0}{2 \sqrt{2\pi} m_e \omega_0} \sqrt{r} \sin{\phi} \cos{(2 \frac{\sqrt{r}}{\sqrt{1-r}} \sin{\phi} \frac{\omega}{\omega_0})} \mbox{sinc}(\frac{N \pi \omega}{2 \omega_0}).
\end{equation}
A similar computation enables us to evaluate the THz spectrum for CP-S pulses as 
\begin{equation}
\label{SpCPS}
{\widehat {\vec E}}^J_{\rm CP-S}(\omega) = \frac{g e^2 N \delta N E_0}{2 \sqrt{\pi} m_e \omega_0} (\sqrt{1-r} +  \frac{\sqrt{r}}{2}) \mbox{sinc}(\frac{N \pi \omega}{\omega_0}) \left( \begin{array}{c} \sin{\phi} \\ \cos{\phi} \end{array} \right).
\end{equation}
From the above expressions, the gain factor at small frequencies $\omega \ll \omega_0$ obtained for CP-S pulses compared to LP-P ones is $G = \frac{\sqrt{2}}{3} (\sqrt{1-r}/\sqrt{r} + 1/2) \approx 1.65$ for $r = 0.1$, i.e., this amplitude gain is close to 2, justifying an increase in the THz power of $\sim 4$.\\

However, the way the THz spectrum builds up for two-color CP-C pulses drastically changes as ${\vec v}_f(t_n)$ now depends on $n$ and leads in the limit $\omega/\omega_0 \rightarrow 0$ to 
\begin{equation}
\label{SpCPC}
{\widehat {\vec E}}^J_{\rm CP-C} (\omega) \approx \frac{g e^2 \delta N E_0}{2 \sqrt{\pi} m_e \omega_0}\left (\sqrt{1-r} - \scriptstyle{\frac{\sqrt{r}}{2}} \right) \left(\cos{\scriptstyle{\frac{N \pi}{3}}} + {\scriptstyle\frac{1}{\sqrt{3}}} \sin {\scriptstyle{\frac{N \pi}{3}}}  \right) \left(\! \begin{array}{>{\scriptstyle}c} \sin{\frac{\phi}{3}} \\ \cos{\frac{\phi}{3}} \end{array}\! \right).
\end{equation}
The THz yield for CP-C beams is again found to be invariant with the two-color relative phase. Also, whereas this yield directly increases for CP-S pulses from the number of ionization events $N$, it evolves like $F(N) \equiv \cos{(N \pi/3)}+ \sin{(N\pi/3)}/\sqrt{3}$ for CP-C ones, which means that their THz field should vanish for pumps containing more than $\sim 3$ optical cycles. This peculiar property holds due the perfect periodicity of the CP-C field length in $3\omega_0$.

\begin{figure}
\centering \includegraphics[width=\columnwidth]{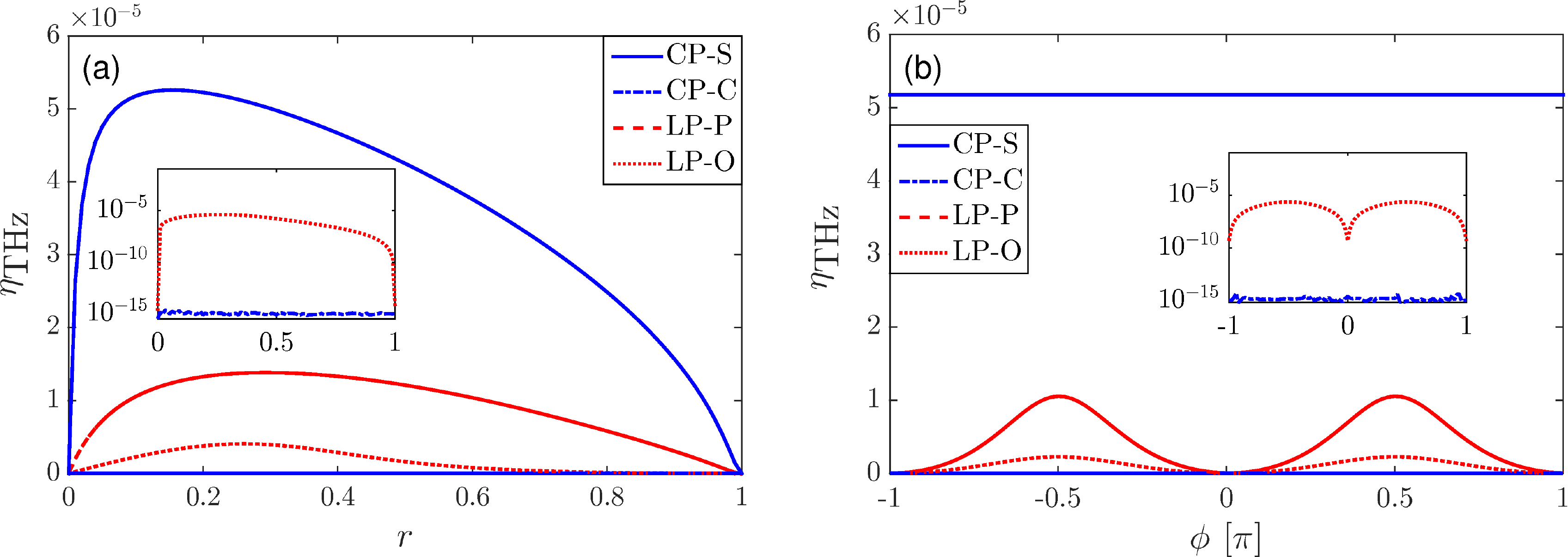}
\caption{(a) Conversion efficiency using a normalization factor in $N_e^{\rm max}$ in the cases LP-P, CP-S and CP-C for 60-fs, 200 TW/cm$^2$ pulses in argon (800-nm FH). (b) Same quantity as a function of the relative phase $\phi$. Insets zoom in the differences in the THz yields supplied by CP-C and LP-O pulses.} 
\label{Fig3}
\end{figure}

The validity of the previous LC evaluations has been checked in the Appendix (see Fig. \ref{Fig2}), in particular their agreement with the experimental behaviors of Ref. \cite{Meng:apl:109:131105}. To end this theoretical investigation, Figure \ref{Fig3} plots the conversion efficiency, $\eta_{THz}$, as a function of the SH intensity fraction $r$ for 200 TW/cm$^2$, 60-fs two-color Gaussian pulses, demonstrating that the enhancement in the THz field produced by CP-S pulses does depend on this fraction. Figure \ref{Fig3}(b) plots the same quantity depending now on the relative phase $\phi$ between the two colors. One sees, in agreement with the vectorial dependencies (\ref{PCCP}) and (\ref{SpCPS}), that the THz yields do not vary with $\phi$. The insets in Fig. \ref{Fig3} detail the weak efficiency achieved by the CP-C and LP-O pulses. Note that their respective order of magnitude is reverted compared to the THz powers experimentally reported by Meng et al. \cite{Meng:apl:109:131105}. This difference may be attributed to imperfect CP polarizations (due to, e.g., a slight ellipticity in the experiments). Indeed, a very small departure from an ideal CP-C configuration may lead to a substantial increase in the THz energy, as clearly illustrated by Fig. \ref{FigAlex2}(b). Note also the sharp slope occurring in the limit $r \rightarrow 0$ for CP pulses, which we relate to the fact that their THz emission is conditioned by the existence of density steps induced by harmonic variations in the ionization rate (see related discussion in Appendix).

\section{3D UPPE simulations of gas jet experiments}
\label{sec3}

The previous properties are now checked by direct 3D numerical computations based on a vectorial version of the unidirectional pulse propagation equation (UPPE)~\cite{Kolesik:prl:89:283902,Kolesik:pre:70:036604} that governs the forward-propagating transverse electric field components $E_x$, $E_y$ of elliptically-polarized pulses: 
\begin{equation}
 \partial _{z}\left( \! \begin{array}{c} \hat{E}_x \\ \hat{E}_y \end{array} \! \right)= {\rm i} \sqrt{k^{2}(\omega)-k_{x}^{2}-k_{y}^{2}}\, \left(\! \begin{array}{c} \hat{E}_x \\ \hat{E}_y \end{array} \! \right) + {\rm i} \frac{\mu _{0}\omega ^{2}}{2k(\omega )}\left( \! \begin{array}{c} \hat{\mathcal{F}}_x^{\mathrm{NL}} \\ \hat{\mathcal{F}}_y^{\mathrm{NL}} \end{array} \! \right), 
\label{1}
\end{equation}
where $\hat{\vec{E}}(k_{x},k_{y},z,\omega )$ is the Fourier transform of the transverse laser electric field components with respect to $x$,
$y$, and~$t$. As in the previous analysis, the longitudinal laser electric field component $E_z$ is neglected. The first term on the right-hand side of Eq.~(\ref{1})
describes linear dispersion and diffraction of the pulse. The
term ${\hat{\vec{\mathcal{F}}}}$$^{\mathrm{NL}}=\hat{\vec{P}}_{\rm Kerr}+ {\rm i} \hat{\vec{J}}/\omega + {\rm i} \hat{\vec{J}}_{\mathrm{loss}}/\omega$ contains the third-order nonlinear polarization $\vec{P}_{\rm Kerr}$ given by Eq.~(\ref{PKerr}) with Kerr index
$n_{2}=3\chi^{(3)}/4n_0^2c\epsilon_0$ $[n_0 = n(\omega_0)]$, the electron current $\vec{J}$ according to Eq.~(\ref{J}), and a loss
term $\vec{J}_{\mathrm{loss}} = [W(E) (N_a-N_e) U_i /E^2 ]\vec{E} $ due to ionization \cite{Berge:rpp:70:1633,Berge:prl:100:113902,Babushkin:prl:105:053903}. We shall first validate our theoretical expectations using the simple QST model for single-ionized argon (ionization potential $U_i = 15.8$ eV) in the intensity range $\sim 200$ TW/cm$^2$, in accordance with the laser parameters chosen in the preceding section.

\begin{figure}[ht]
\centering \includegraphics[width=\columnwidth]{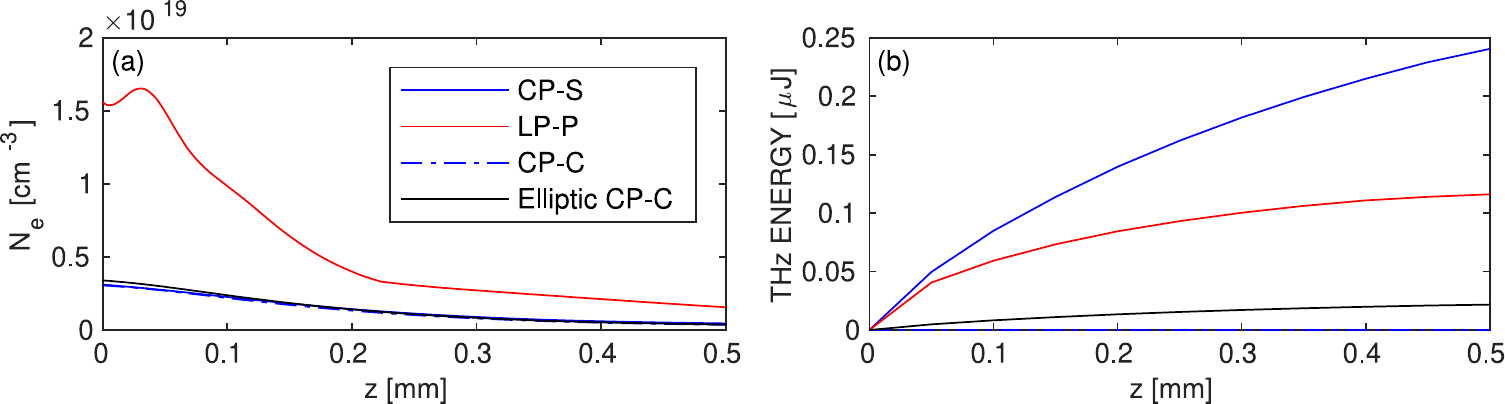}
\caption{3D UPPE simulations of two-color Gaussian beams with different polarization states propagating over 500 $\mu$m in a gas jet configuration (argon) for different polarization states: LP-P (red solid curve), CP-S (blue solid curve), and CP-C pulses with 60-fs FWHM duration (dash-dotted blue curve). The black curve refers to an elliptically-polarized pulse close to a CP-C pump with $\rho =0.9$ and $\theta = - 0.446 \pi$. (a) Peak electron density and (b) THz energy yields along the propagation axis corresponding to the above plotstyles.}
\label{Fig6}
\end{figure}

We here simulate gas jet experiments, i.e., setups using micrometer-sized pulse beams propagating over short ranges, $z \leq 500\,\mu$m, in order to limit the linear (diffraction, dispersion) and nonlinear (Kerr response, plasma generation, energy loss) propagation effects that can affect the laser pump components and induce strong variations in their relative phase. Over such short optical paths the pulse intensity (not shown) does not experience strong variations. With an 800-nm FH pump, the two colors have equal input beam width $w_0 = 50\,\mu$m at 1/e$^2$ intensity and FWHM duration of 60 fs. Their peak power is subcritical, $P_{\rm in} = 0.77\,P_{\rm cr}$, with $P_{\rm cr} \simeq \lambda_0^2/2\pi n_2 = 10.2$ GW and $n_2 = 10^{-19}$ cm$^2$/W following \cite{Loriot:oe:17:13429} for argon at atmospheric pressure. Simulations have been performed using a time window of 0.8 ps, a temporal step of $\Delta t = 50$~attoseconds and transverse resolution of $\Delta x = \Delta y \approx 0.78$ $\mu$m. 

Figure \ref{Fig6} displays the electron density and THz energy yield extracted in the frequency window $\nu \leq \nu_{co} = 90$ THz for different two-color pump arrangements in LP-P (red solid curve), CP-S and CP-C configurations (solid and dash-dotted blue curves, respectively). The input relative phase between the two colors is set equal to $\pi/2$ and the SH intensity fraction is $r = 10\%$. As seen from Fig. \ref{Fig6}(a), whereas the density created by the CP pulses experiences a smooth attenuation along propagation, that generated by the LP-P pulse undergoes a sharper decrease due to the higher ionization yield enhancing plasma losses. Figure \ref{Fig6}(b) reveals the gain in the THz energy yield between CP-S and LP-P reaching a factor $\sim 1.44$ at $z = 100\,\mu$m for an electron density ratio between the two pulse configurations of $N_e^{\rm LP-P}/N_e^{\rm CP-S} \approx 5$. This gain is even amplified to 2.1 for a density ratio reduced to $N_e^{\rm LP-P}/N_e^{\rm CP-S} \approx 3.76$ after 500 $\mu$m of propagation. Thus, the THz relative gain reported to the same ionization yield is about 7.2 - 7.9, which reasonably agrees with our LC computations (factor $\sim 6$, see Figs. \ref{FigAlex2} and \ref{Fig3}). This gain is all the more important as the electron density decreases in the LP-P configuration and the corresponding THz emission saturates. In addition the black curve of Fig. \ref{Fig6} represents the same propagation features for a pump pulse consisting of a perturbed CP-C state with an ellipse ratio reduced by $10\,\%$ and a relative phase slightly shifted from $\pi/2$ with $\theta = - 0.446 \pi$. One can observe that perturbing a CP-C state fosters a significant THz emission, which is in agreement with Fig. \ref{FigAlex2}(b).

\begin{figure}
\includegraphics[width=\columnwidth]{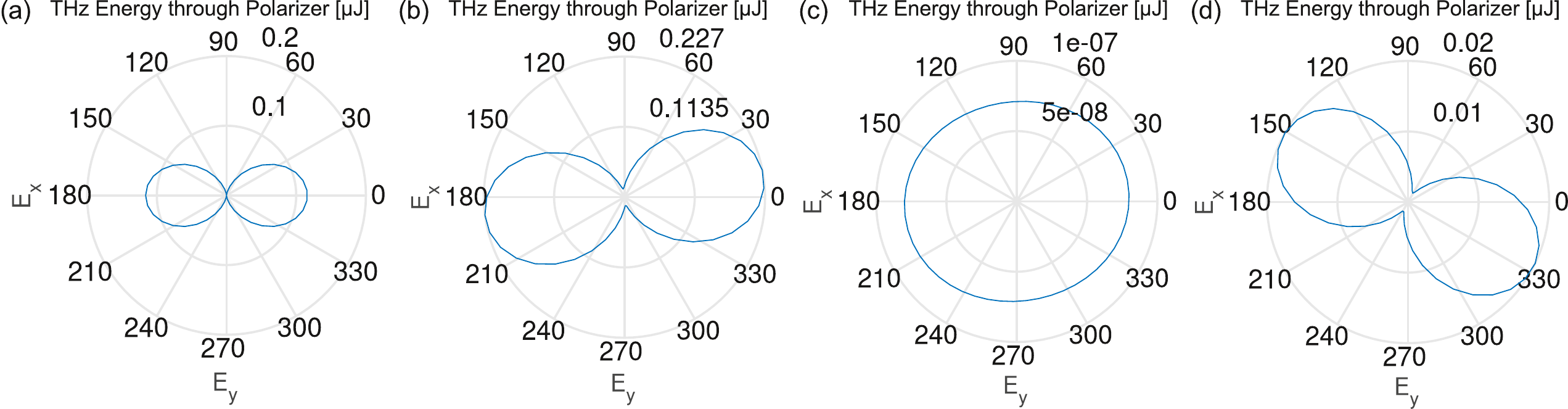}
\caption{THz transmitted energy through a polarizer obtained at $z=500\,\mu$m from 3D UPPE simulations of two-color Gaussian beams in an argon gas jet for different polarization states: (a) LP-P, (b) CP-S, (c) CP-C, and (d) an elliptically-polarized pulse close to CP-C configuration with an ellipse ratio decreased by 10 $\%$ and $\theta = - 0.446 \pi$. The numbers refer to iso-contour energy levels.}
\label{Fig7}
\end{figure}

Looking further into the THz pulse structure, Fig. \ref{Fig7} details the THz polarization patterns for the three baseline simulations with $\tau = 60$ fs. As can be seen from Figs. \ref{Fig7}(a,b), the CP-S pulse produces a THz polarization pattern being slightly tilted, although the emitted THz field remains mainly polarized along the $x$ axis, as expected from Eqs. (\ref{PCLP}), (\ref{PCCP}) and (\ref{SpCPS}) when $\phi = \pi/2$. By contrast, the THz pulse produced by the CP-C pulse does not exceed the background noise level, i.e., no relevant THz waveform is generated, which justifies its isotropic polarization pattern [Fig. \ref{Fig7}(c)]. Figure \ref{Fig7}(c) displays the same information for the elliptically polarized pulse close to a CP-C state, but with an ellipse aspect ratio diminished by $10\%$ and $\theta = - 0.446 \pi$. We checked that for CP pumps, the polarization ellipse of the emitted THz field rotated accordingly with the input value of the relative phase $\phi$, in accordance with our analytical formulas. 

\section{Experimental setup and results}
\label{sec4}

\begin{figure}
\centering 
\includegraphics[width=\columnwidth]{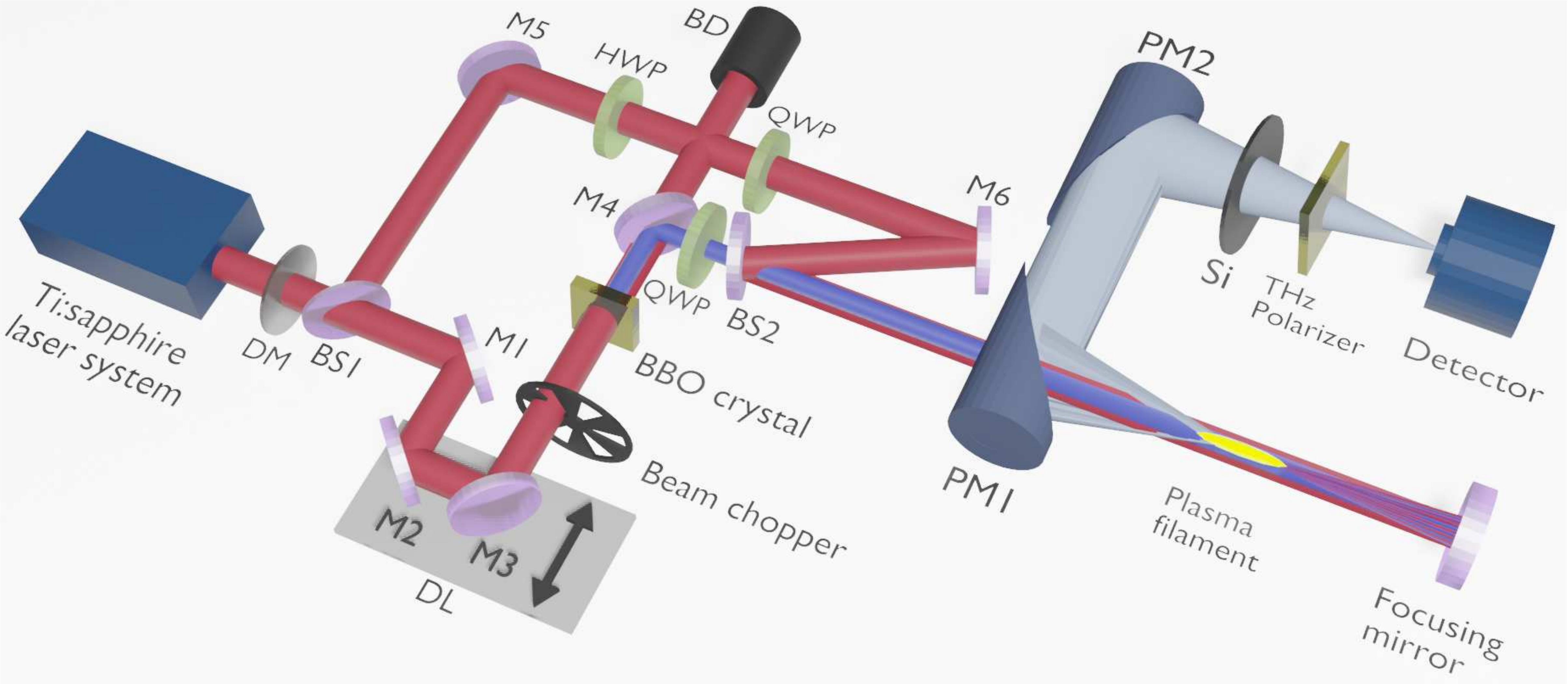}
\caption{Experimental setup. DM, M1-M6: dielectric mirrors; HWP and QWP: half- and quarter-wave plates;  BS1, BS2: beam splitters, PM1, PM2: parabolic off-axis mirrors; DL: optical delay line; BD: beam dump. SH and FH are shown by blue and red-pink color, respectively. THz radiation is represented by gray color.}
\label{FigExp1}
\end{figure}

Terahertz waves were generated in air by bichromatic femtosecond laser pulses following the setup shown in Fig. \ref{FigExp1}.  A 1 kHz repetition rate femtosecond Ti:sapphire chirped pulse amplification laser system (Legend elite duo HE+, Coherent Inc.) delivering 40-45 fs (FWHM) light pulses centered at 790 nm with maximal pulse energy of 8 mJ was used as a pump source. The output laser power could be varied by inserting thin partially reflecting dielectric mirrors (DMs) into the beam path. The laser beam was divided into two arms thanks to a thin 50:50 beam splitter (BS1). One of these beams was used for second harmonic generation through a 0.2 mm thick nonlinear BBO crystal. A temporal delay between the fundamental harmonic (FH) and second harmonic (SH) pulses was introduced by using a motorized optical delay line (DL). The pulse polarization was controlled using broadband zero-order half- and quarter-wave plates (HWP and QWP, respectively) inserted into the beam paths. The QWPs allowed to vary the polarization state of both FH and SH pulses from linear to circular, while HWP inserted into the FH beam path alone was used to switch between the mutually orthogonal and parallel linear polarizations. The FH and SH beams were concentrically superimposed at a dichroic beam splitter (BS2). After passing through the hole of an aluminum-coated off-axis parabolic mirror (PM1) they were directed to a focusing spherical mirror (focal length about 22 cm). As a result, a visible few-mm long plasma filament was produced. In order to minimize optical aberrations the focused bichromatic pump beam was reflected nearly exactly in the backward direction by the focusing mirror. Despite its hole at center, the mirror PM1 was still capable to collect and collimate most of emerging THz radiation, which formed a hollow cone with $\sim 5^\circ$ apex angle \cite{Zhong:apl:88:261103,Vaicaitis:lp:28:095402}. The second parabolic mirror (PM2) then focused the THz beam onto the pyroelectric detector (TPR-A-65 THz, Spectrum Detector Inc.), sensitive in the range 0.1 - 300 THz (3000 - 1 $\mu$m) with a flat response function from $\sim 3$ to $\sim 100$ THz. From FH and SH waves THz radiation was separated by a few 0.5-1 mm-thick high-resistivity silicon wafers. The polarization state of the detected THz beam could be monitored by a broadband HDPE THz polarizer (Tydex Co.), placed in front of the pyroelectric detector. In order to enhance the detector sensitivity a lock-in amplifier (SR530, Stanford research systems) along with an optical beam chopper placed into the SH beam path was utilized. In order to reduce fluctuations of the THz signal the lock-in time constant was kept to be 300 ms in most cases, while the repetition rate of the beam chopper was 5 Hz. Data acquisition and processing was performed by means of a computer and appropriate software, which also controlled parameters of the optical delay line and THz polarizer. Imaging the generated THz beam was performed with a thermal camera detector (VarioCAM head HiRes 640, InfraTec GmbH), sensitive in the range 0.1- 40 THz.

Note that, although our experimental setup allowed for an independent control of the intensity, polarization and transverse positions of the focused FH and SH beams, the relative phase between the latter could not be monitored during the experiment, as the system was not interferometrically stabilized. Therefore, the lack of phase stability sometimes resulted in strong fluctuations in the measured THz yields, preventing us from further investigating the influence of the two-color relative phase on the THz generation process. However, our diagnostics were able to average the THz signal over time and consequently, over the relative phase between the FH and SH waves, which allowed us to investigate the main properties of the generated THz radiation as a function of the pump polarization states.

\begin{figure}
\centering 
\includegraphics[width=0.75\columnwidth]{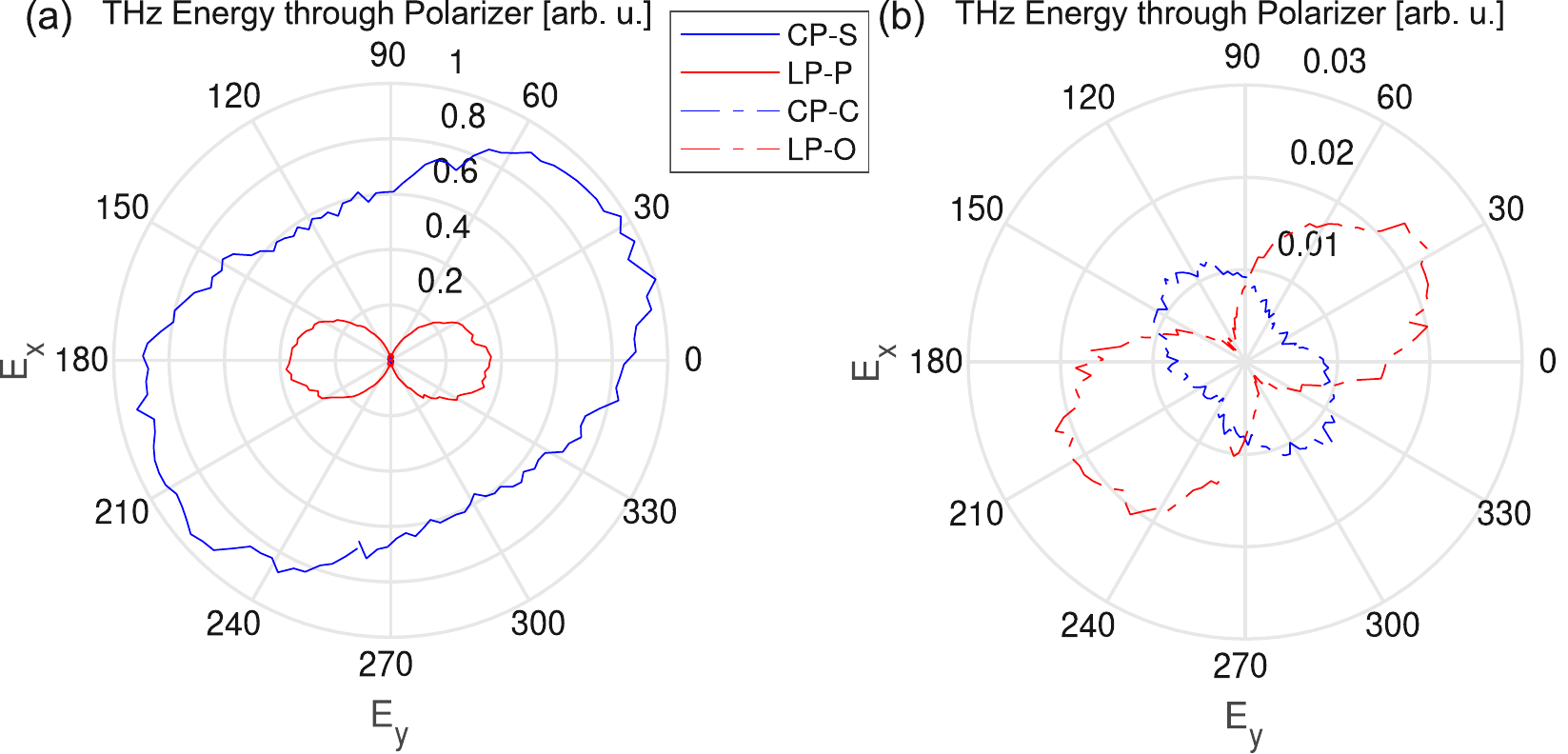}
\caption{Experimental THz yield versus THz polarizer angle in arbitrary but comparable units. (a) CP-S (solid blue line) and LP-P (solid red line). (b) CP-C (dashed dotted blue line) and LP-O (dashed dotted red line). Energies of the FH and SH pulses were 0.92 and 0.03 mJ, respectively ($r \approx 3.3\%$).}
\label{FigExp2}
\end{figure}

Figure \ref{FigExp2} displays the THz yield measured as a function of the THz polarizer (placed in front of the detector) angle for the CP-S, CP-C, LP-P and LP-O pulses with FH and SH pulse energies of 0.92 and 0.03 mJ, respectively. One can see that, in line with our theoretical predictions (see Fig \ref{Fig3}), the THz yield is the strongest in the CP-S case. A significantly lower THz signal has been obtained from linearly polarized pump pulses (lower by a factor of 4.3 and 67 for LP-P and LP-O configurations, respectively). As predicted, the lowest THz yield is obtained for circularly polarized counterrotating (CP-C) pump waves. Note that, though the numerical analysis indicated that in the CP-C case THz generation could be weaker by about seven orders of magnitude compared with that obtained using LP-O pulses, our experimental THz yield was lower approximately only by a factor $\sim 1.4$. This difference can be explained by the fact that it is experimentally quite difficult to produce ideal CP-C pumps, which may significantly rise the efficiency of THz generation. In Meng et al.'s experiments \cite{Meng:apl:109:131105} the measured THz yield in CP-C case was even higher than that obtained in the LP-O configuration, which could also be explained by non-ideal circularities of the pump polarization states. The fact that slightly perturbed CP-C pulses are able to create a significant THz yield highlights the very peculiar property of ideal CP-C pulses in ''killing'' THz waves, as justified in Sec. \ref{sec2}.

Figure \ref{FigExp2} also contains partial information on the generated THz  polarization state. As expected, for the LP-P pulses THz polarization is clearly linear. An almost linear THz polarization state was also registered for the LP-O pulses. However, from LC theory one would expect the THz radiation to be polarized like the SH pulse ($x$-polarized). This is obviously not the case, which we attribute to slightly non-orthogonal FH and SH polarization in our experiment. In agreement with observations reported in~\cite{Kosareva:ol:43:90}, the LC model predicts that THz polarization is very sensitive to the pump polarization. For example, a deviation of only 5 degrees from the ideal 90 degrees is enough to turn the THz polarization by 20 degrees, and sub-optimal phase angle $\phi$ may render THz polarization even elliptical.

\begin{figure}
\centering \includegraphics[width=0.45\columnwidth]{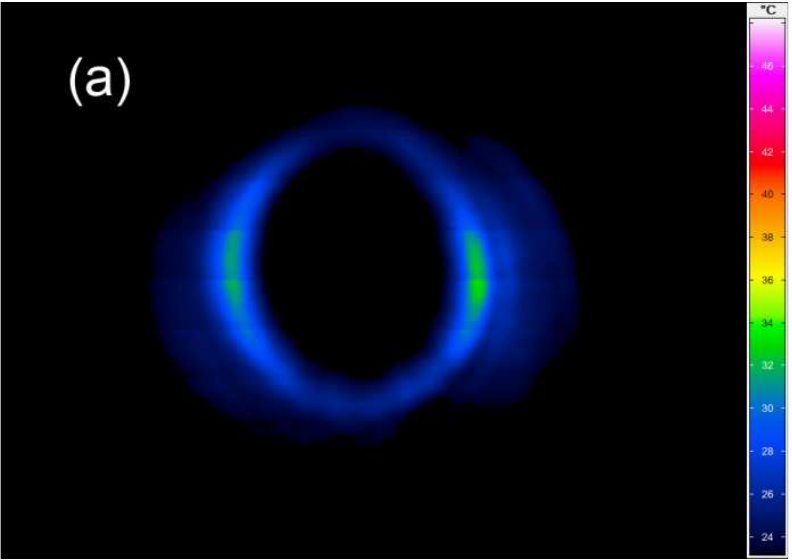}
\includegraphics[width=0.45\columnwidth]{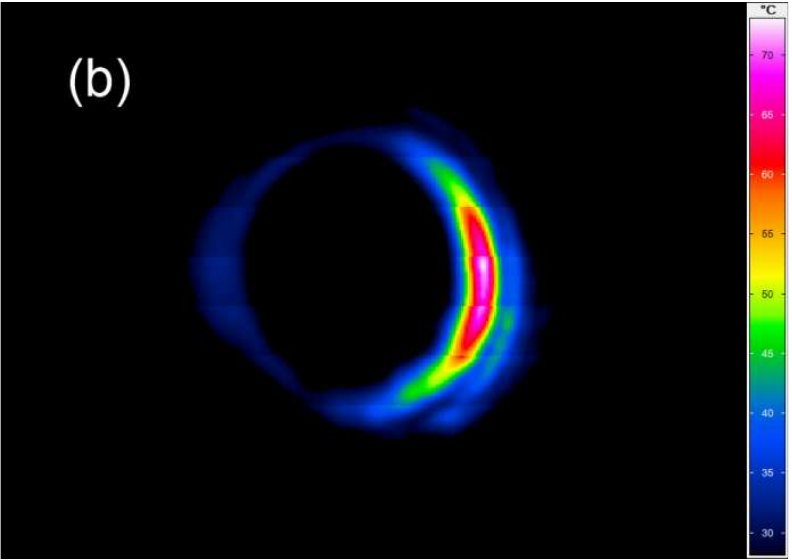}
\caption{THz far-field beam patterns registered (a) at almost zero mutual FH and SH beam displacement and tilt and (b) for the beam displacement of about 40 $\mu$m and tilt of about $-0.7$ mrad.}
\label{FigExp4}
\end{figure}

For the CP-S configuration, in contrast to our theoretical predictions, there was only a rather weak dependence in the registered THz signal on the polarizer angle. This rather surprising result could indicate elliptical THz polarization. However, as explained in the next section, this property is caused by imperfect transverse alignment of the two pump colors, in such a way that no THz polarization state can be defined.
A larger transverse pump beam displacement and tilt in the focal plane could also significantly modify the spatial intensity distribution of the generated THz radiation. When the beam displacement and tilt were nearly zero, a standard hollow THz cone was produced [see Fig. \ref{FigExp4}(a)]. Nonetheless, even slight deviations altering the on-axis, collinear propagation of the two colors resulted in the redistribution of the THz intensity to one side of the beam, creating typically a ''young moon'' pattern [see Fig. \ref{FigExp4}(b)].  

\section{Role of the transverse shifts in the color alignement}
\label{sec5}

In this final section, the above experimental setup is numerically simulated by means of our 3D UPPE vectorial model. Because experiments were performed in ambient air, we use the effective nonlinear refractive index $n_2=1.3 \times 10^{-19}$~cm$^2$/W, mimicking the joint contribution of instantaneous and Raman delayed responses~\cite{Wahlstrand:pra:85:043820,Rosenthal:jpb:48:094011} for the sub-50-fs pump-pulses employed here. Linear dispersion of air was taken from~\cite{Peck:josa:62:958}. Ionization of both oxygen and nitrogen molecules was taken into account, by employing field-dependent PPT rates for both species and adopting Talebpour et al.'s charge numbers $Z^*_{O_2}=0.53$, $Z^*_{N_2}=0.9$~\cite{Talebpour:oc:163:29}.

\begin{figure}
\centerline{\includegraphics[width=\columnwidth]{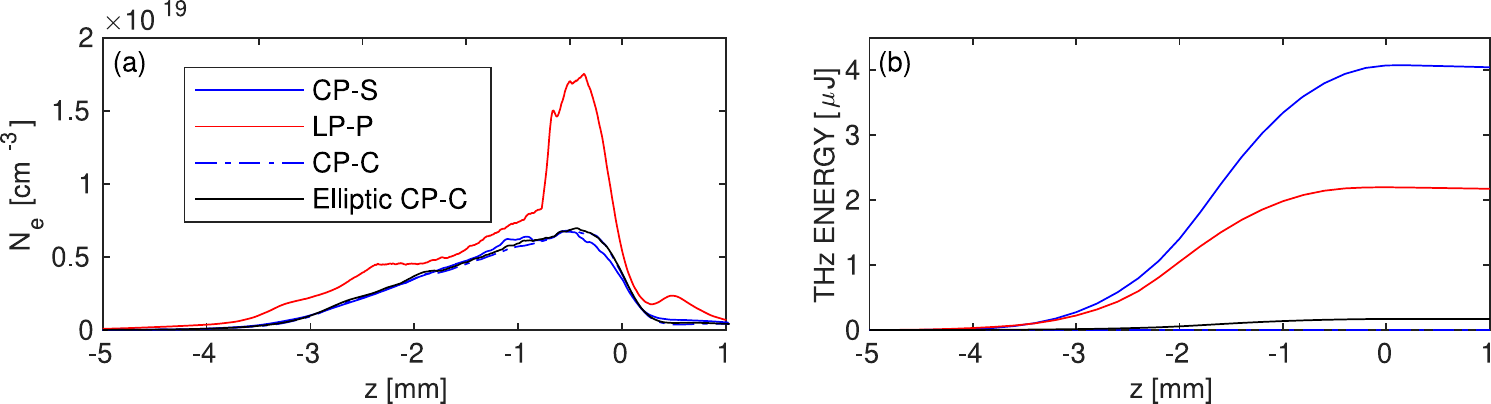}}
\caption{3D UPPE simulations of two-color Gaussian beams corresponding to the above experiments in air (Fig.~\ref{FigExp2}) for different polarization states: LP-P (red solid curve), CP-S (blue solid curve), and CP-C (dash-dotted blue curve). The black curve refers to an elliptically-polarized pulse close to a CP-C pump with $\rho =0.9$ and $\theta = - 0.446 \pi$. (a) Peak electron density and (b) corresponding THz energy yields along the propagation axis. Note that the electron density triggered by CP-S and CP-C pulses almost overlap while the CP-C pulse-driven THz energy is close to zero.}
\label{Fig13}
\end{figure}

Figure~\ref{Fig13} shows results for various pump polarization states and perfect beam alignment. The simulated peak plasma densities in Fig.~\ref{Fig13}(a) are comparable for all cases and confirm that a few mm long plasma is formed before the geometrical focus (here located at $z=0$). As expected, LP-P pump-polarization produces more plasma than CP beams for the same pulse energies and durations. The THz energy produced in the plasma in Fig.~\ref{Fig13}(b) confirms once more the impact of the pump polarization as discussed in the previous sections. In order to get significant THz yield from CP-C configuration, a slight ellipticity is required. 

\begin{figure}
\includegraphics[width=\columnwidth]{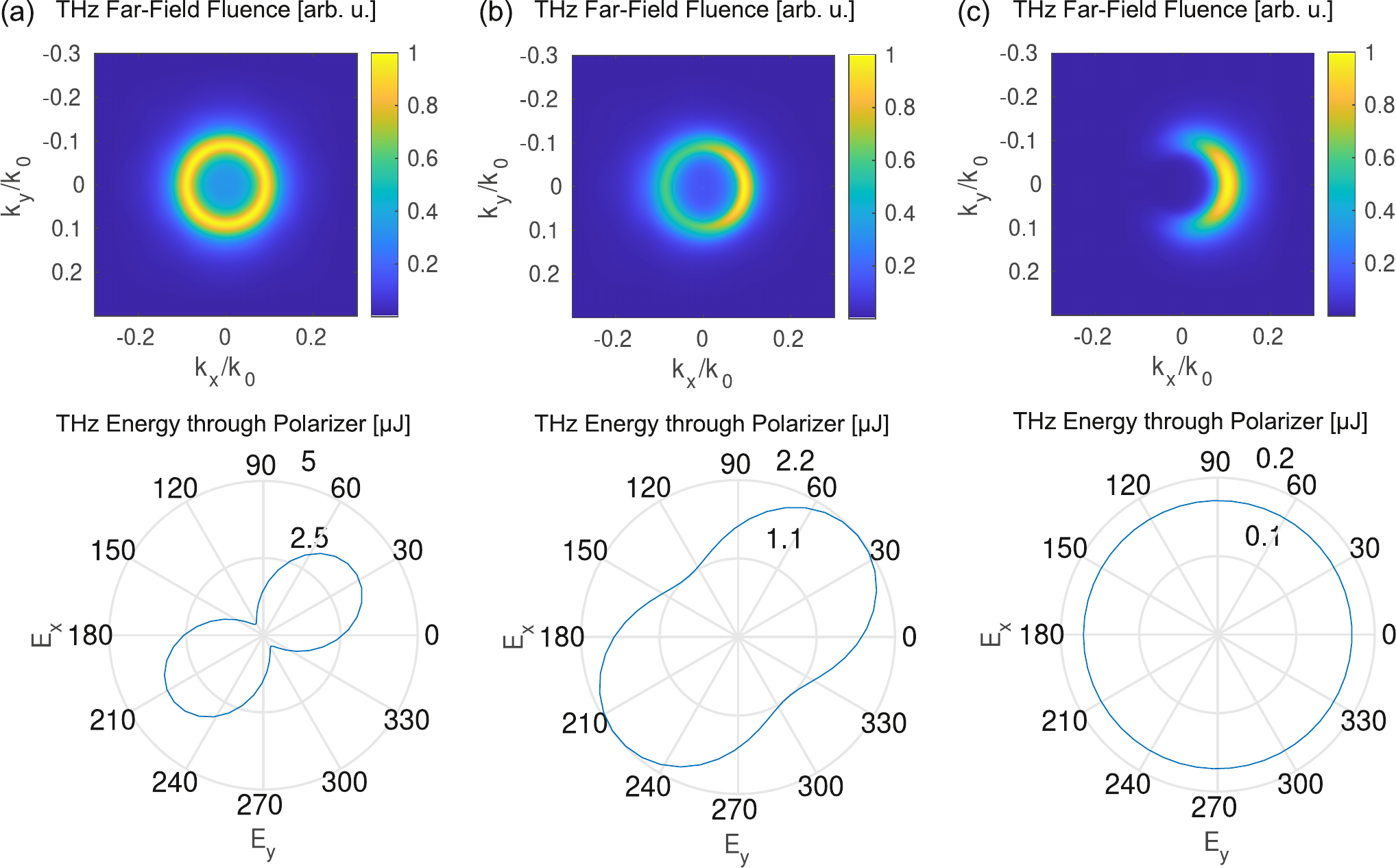}
\caption{CP-S THz far-field fluence ($\nu<90$~THz). (a) Perfect beam alignment. (b) SH beam displacement of about 4~$\mu$m and tilt of about $-0.04$~mrad. (c) SH beam displacement of about 40~$\mu$m and tilt of about $-0.4$~mrad.}
\label{Fig14}
\end{figure}

Let us now have a closer look at the CP-S case, with particular attention paid to the impact of transverse beam displacements. Figure~\ref{Fig14}(a) shows the THz far-field fluence pattern in the frequency window $\nu<90$~THz for perfect beam alignment. As in the experiment, conical THz emission leads to a ring pattern. The THz energy versus polarization indicates almost linear polarization, close to what was observed in the gas-jet simulations of Section~\ref{sec3}. Due to our two-arm experimental setup, the transverse beam alignment in the focal plane is an issue. In our experimental setup, transverse beam displacement is linked to a tilt angle, which we estimate to $\sim -0.01$~mrad per $\mu$m of transverse displacement. This tilt angle can be controlled with a precision of about $0.1$~mrad, which translates to an uncertainty of about $\pm~5$~$\mu$m in the transverse relative beam position between FH and SH at focus. Figure~\ref{Fig14}(b) shows the resulting impact on the far-field fluence pattern for such \emph{small} displacement computed with the UPPE model. The ring pattern is still visible, and compares well to the experimental pattern in Fig.~\ref{FigExp4}(a). Most importantly, the THz energy distribution versus polarization now agrees with that shown in Fig.~\ref{FigExp2}(a). This means that a small transverse misalignment of SH and FH, consistent with the limits of our setup, can explain the seemingly elliptical THz polarization observed in our experiments. 

On the other hand, \emph{larger} displacements between FH and SH can be introduced deliberately. Such a configuration is shown in Fig.~\ref{Fig14}(c), where the transverse displacement is increased by one order of magnitude compared to Fig.~\ref{Fig14}(b). A ''young moon'' pattern is then obtained, comparable to the one shown in Fig.~\ref{FigExp4}(b). This pattern occurs on the same side to which the SH beam is displaced and directly results from breaking the rotational symmetry. For such large beam displacement, the THz polarization looks completely ''circular'' - or isotropic. We here want to stress that the resulting THz radiation does not have in fact a well-defined polarization state. It just proceeds from a superimposition of THz waves with different polarization states originating from different locations in the plasma, causing the THz energy transmitted through a polarizer to be invariant to the polarizer angle.

\begin{figure}
\includegraphics[width=\columnwidth]{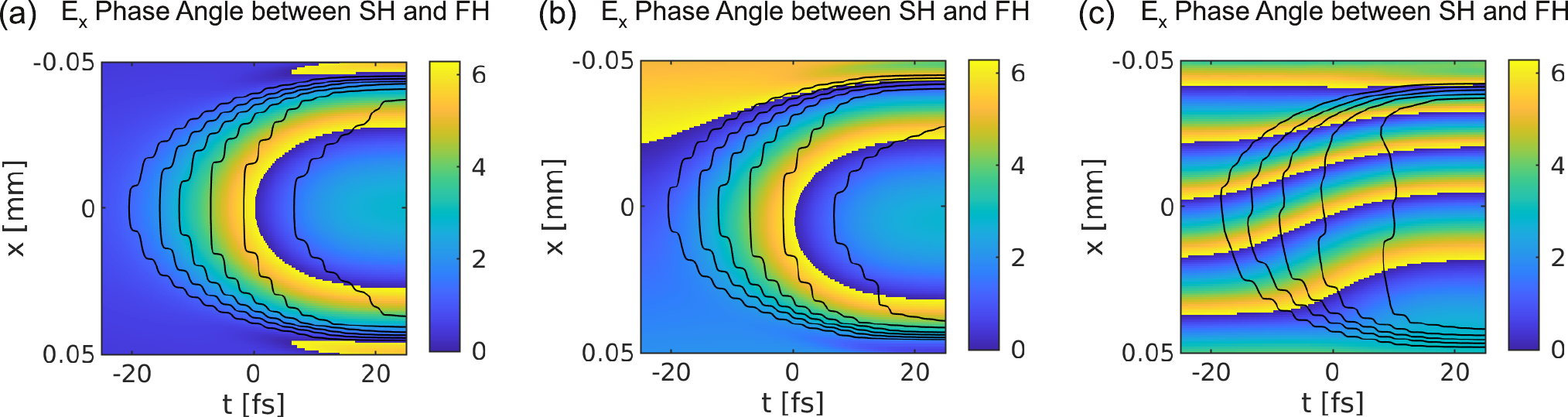}
\caption{Phase angle between $E_x^{\rm SH}$ and $E_x^{\rm FH}$ [$\arg{(E_x^{\rm SH})}-2\arg{(E_x^{\rm FH})}$] for CP-S pump pulses in the $y=0$ plane, 2~mm before the geometrical focus for (a) perfect beam alignment, (b) beam displacement of about 4~$\mu$m and tilt of about $-0.04$~mrad, and (c) beam displacement of about 40~$\mu$m and tilt of about $-0.4$~mrad. The black lines show iso-electron density surfaces.}
\label{Fig15}
\end{figure}

As explained in Section~\ref{sec2}, in CP-S configuration the phase angle $\phi$ between SH and FH determines the orientation of the (quasi-linearly) polarized THz radiation. Therefore, it is instructive to plot this quantity at a position of strong THz generation inside the plasma. Figure~\ref{Fig15} shows the relative phase between the FH and SH components in the plane $y=0$ for the three pump configurations of Fig.~\ref{Fig14}. The first important information is that the relative phase is not constant, but a function of the transverse spatial coordinates and time. Yet, THz radiation can be produced with different polarization states depending on the ionization response along propagation which is directly conditioned by the relative phase locally achieved by the two pump harmonics. Superimposing iso-electron-density lines to the phase plots can thus be instructive. For the perfectly aligned beams in Fig.~\ref{Fig15}(a), these iso-electron-density lines follow more or less the phase landscape, so that THz generation favors the standard linear polarization orientation expected in this case, as observed in Fig.~\ref{Fig14}(a). However, already a \emph{small} misalignment of FH and SH beams makes  iso-electron-density lines cross regions with significantly varying relative phase, in particular when comparing locations above and below the optical axis. The polarization of the THz waves produced above and below the optical axis differs, which results in an ''elliptic'' THz polarization shown in  Fig.~\ref{Fig14}(b). Again, it is important to keep in mind that the resulting THz radiation is \emph{not} strictly speaking elliptically polarized, but it results from a complex spatio-temporal pattern that, once averaged, supplies partially polarized THz light. This assessment is further corroborated by Fig.~\ref{Fig15}(c), where due to large beam displacement the iso-electron-density lines and relative phase landscape are completely disconnected. As a result, no preferred direction of polarization is detected by the polarizer in Fig.~\ref{Fig14}(c).

We note that other effects may affect the relative phase between FH and SH along the mm-long plasmas and make the THz polarization change along $z$. Among those effects are the Gouy phases of SH and FH beams, plasma dispersion, as well as slightly different focus positions for SH and FH. Our simulations indicate, however, that the impact of all these effects on the THz polarization is much smaller than that of the above-discussed transverse beam displacements.

\section{Conclusion}
\label{sec6}

In this paper, we have investigated the efficiency of the photocurrent mechanism to produce THz pulses when circularly- and elliptically-polarized two color laser pulses are employed. Theoretical evaluations based on plane wave and local current analyses displayed evidence that a direct dependency of the drift velocity acquired by photo-ionized electrons on the FH pump amplitude and longer ionization sequences explain the increase in the THz power reported by recent experimental and numerical investigations exploiting circularly-polarized two color pulses. When FH and SH pump components counter-rotate, the complete vanishing of the emitted THz radiation has been shown to proceed from destructive interferences linked to the third-harmonic periodicity of the overall laser electric field. Introducing small ellipticity may, by contrast, allow to recover relevant THz energies. These results have been confirmed by direct full (3D+1) UPPE simulations of short, gas-jet plasmas.

In addition, our experimental measurements using a two-armed setup recovered the main trends on THz generation efficiencies for linearly and circularly polarized pump pulses over longer (mm-long) plasma-filament ranges. However, our THz polarization measurements did not confirm the occurrence of linearly-polarized THz fields predicted by our theoretical findings, but instead some undefined polarization state, with either elliptical or even circular energy distribution. Comprehensive 3D simulations revealed that such polarization patterns occur due to slight transverse beam displacements and tilts of the pump harmonics in the focal plane, which causes generation of THz waves with different polarization orientations that superimpose. This insight is of crucial importance for applications sensitive to the THz polarization state. We expect that our results will pave the way towards more performant and stable broadband THz sources, and will trigger future experiments in this field. 

\appendix

\section{Calculation of electron drift velocities and LC results}
 Assuming long enough pulses to treat them with slowly varying envelope, Eq. (\ref{vf}) can be expanded as
\begin{eqnarray}
& {\vec v}_f(t)  \simeq - \frac{e E_0 \mathrm{e}^{-2 \ln 2 \frac{t^2}{\tau^2}}}{m_e \sqrt{1+\rho^2}} \label{vf2}\\
& \times \! \left( \! \begin{array}{c}  \frac{\sqrt{1-r}[\nu_c \cos\!{(\omega_0 t)} + \omega_0 \sin\!{(\omega_0 t)}]}{\nu_c^2 + \omega_0^2} +  \frac{\sqrt{r}[\nu_c \cos\!{(2 \omega_0 t + \phi)} + 2 \omega_0 \sin\!{(2 \omega_0 t + \phi)}]}{\nu_c^2 + 4\omega_0^2}  \\   \frac{\rho\sqrt{1-r}[\nu_c \cos\!{(\omega_0 t + \theta)} + \omega_0 \sin\!{(\omega_0 t +\theta)}]}{\nu_c^2 + \omega_0^2} +   \frac{ \rho \epsilon\sqrt{r}[\nu_c \cos\!{(2 \omega_0 t + \theta + \phi)} + 2 \omega_0 \sin\!{(2\omega_0 t + \theta + \phi)}]}{\nu_c^2 + 4\omega_0^2}    \end{array} \! \right). \nonumber 
\end{eqnarray}
This expression directly provides the drift electron velocities evaluated at $t = t_n$ for a collisionless plasma $(\nu_c \rightarrow 0)$ and assuming $r \ll 1$, namely,
\begin{equation}
\label{vfLPP}
v_f^{\rm LP-P} (t_n) = 3 e E_0 \sqrt{r} \sin{\phi}/(2m_e \omega_0),
\end{equation}
\begin{equation}
\label{vfCPS}
{\vec v}_f^{\rm CP-S} (t_n) = \frac{e E_0}{\sqrt{2} m_e \omega_0} (\sqrt{1-r} + \frac{\sqrt{r}}{2}) \left( \begin{array}{c} \sin{\phi} \\ \cos{\phi} \end{array} \right),
\end{equation}
and
\begin{equation}
\label{vfCPC}
{\vec v}_f^{\rm CP-C}(t_n) = \frac{e E_0}{\sqrt{2} m_e \omega_0} (\sqrt{1-r} - \frac{\sqrt{r}}{2}) \left( \begin{array}{c} \sin{(\phi/3 - 2 n \pi/3)} \\ \cos{(\phi/3 - 2 n \pi/3)} \end{array} \right),
\end{equation}
for the LP-P, CP-S and CP-C pump configurations associated with the ionization instants $\omega_0 t_n^{\rm LP-P} \approx n \pi - 2 \sqrt{r} (-1)^n \sin{\phi}/\sqrt{1-r}$, $\omega_0 t_n^{\rm CP-S} = 2 n \pi - \phi$ and $\omega_0 t_n^{CP-C} = 2n \pi/3 - \phi/3$, respectively. 

The information gained through this microscopic description is the direct dependency of the electron velocity (\ref{vfCPS}) on the dominant FH amplitude and not on the SH one as is the case of LP-P pulses. Note that, for LP-P pulses and CP-S pulses as well, the kick in the electron momenta $\propto v_f(t_n)$ is here independent on the ionization instant $n$, so that the THz spectrum (\ref{Sp}) proceeds from the sum $\sum_n \mbox{e}^{i \omega t_n}$ only.

It is worth emphasizing that, in the present analysis, ${\vec v}_f(t_n)$ does not reduce to zero for CP pulses in the limit of no SH $(r \rightarrow 0)$. In this limit, the emitted THz field should, however, vanish, as shown in Fig. \ref{Fig3}(a). This behavior follows from the fact that the absence of SH color limits the ionization yield to a smooth increase in the electron density that just follows the envelope of the FH pulse. Reversely, our microscopic model is based on a steplike increase in the density, which can only be realized through the existence of short ionization instants compared to the width of the density steps, typically $\tau_n^{\rm ion}\ll \Delta t_n = t_{n+1}-t_n$. A straightforward manipulation of Eq. (\ref{taun}) together with Eq. (\ref{length2}) leads to the diverging behavior $\tau_n^{\rm ion} \simeq 2^{1/4} \sqrt{E_0/\beta}  (\omega_0 r^{1/4})^{-1}$ in the limit of small $r$, so that, e.g., the requirement $\tau_n^{\rm ion} \leq \Delta t_n/10 = \pi/5 \omega_0$ for CP-S pulses supplies the minimum bound $r \geq r_{\rm lim}= 10^4 E_0^2/(8 \pi^4 \beta^2)$. Therefore, for CP pump pulses Eqs. (\ref{SpCPS}) and (\ref{SpCPC}) are only valid for large enough SH intensity fraction $r$. For smaller $r$, and therefore longer ionization events, electrons born at different times than $t_n$ start to contribute to the current. Because their trajectory is different, they do not contribute to the low frequency component emitting the THz radiation; in the worst case they may even produce a current with opposite sign, reducing thereby the radiation yield.

\begin{figure}
\centering \includegraphics[width=\columnwidth]{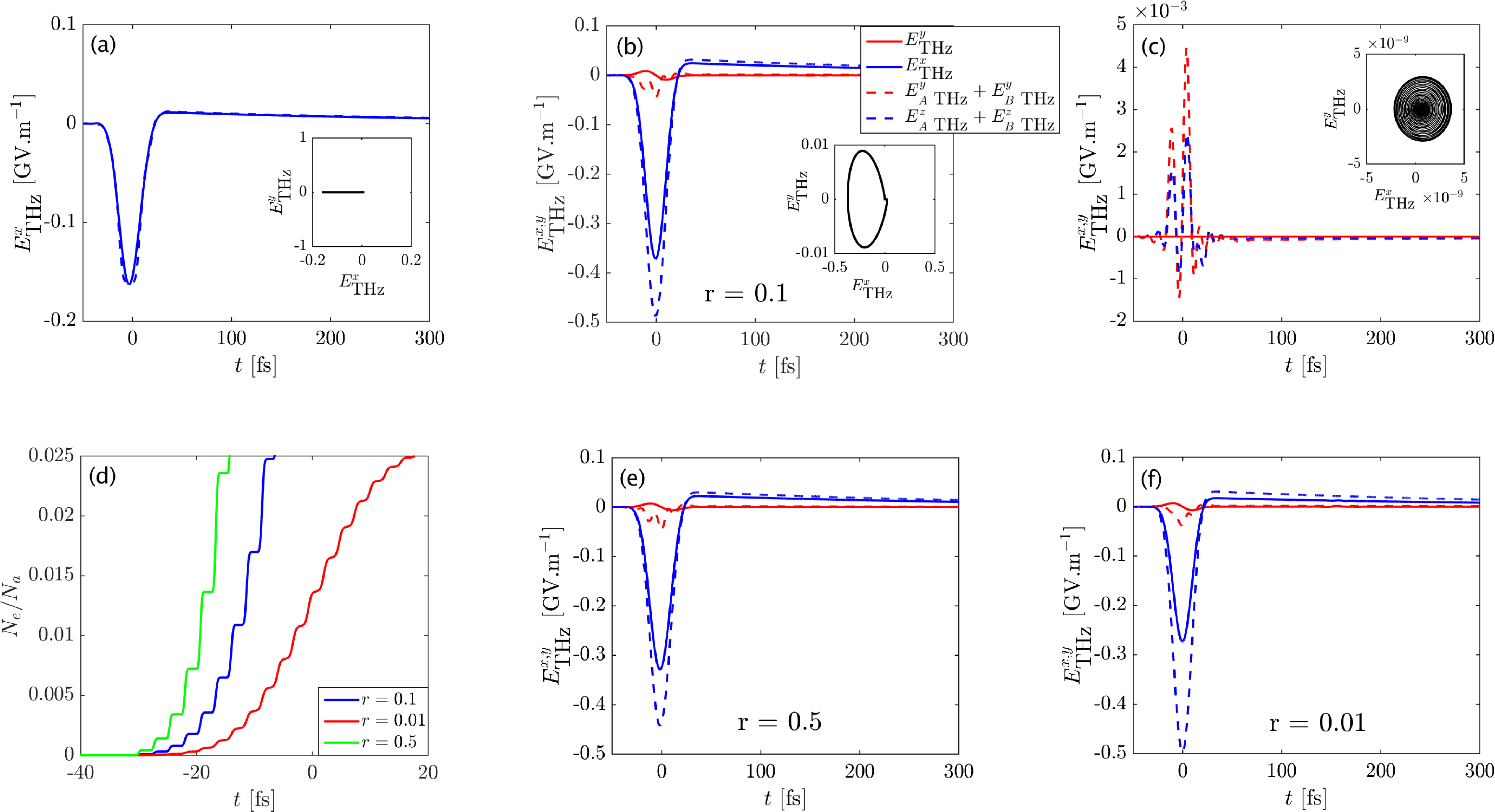}
\caption{(a,b,c) THz waveforms evaluated semi-analytically (dashed curves) and by the LC model integrated numerically (solid curves) in the three configurations of interest (LP-P, CP-S, CP-C) issued from the pump configurations of Fig. \ref{Fig1} for $r = 0.1$. Insets show their corresponding polarization states. (d) Variations in time of $N_e/N_a$ for a CP-S case with different SH intensity fraction $r$. (e,f) Same curves as in (a-c) for (e) $r = 0.5$ and (f) $r = 0.01$.}
\label{Fig2}
\end{figure}

Figures \ref{Fig2}(a,b,c) illustrate the on-axis field components (dashed curves) expected from the current components (\ref{current}) and (\ref{current_2}) for the same Gaussian pulses as in Fig. \ref{Fig3}. Compared with LP-P pulses the THz waveform computed over the frequency range $\nu \leq 90$ THz and generated over the laser region ($|t| \leq 60$ fs) is higher in CP-S configuration by a factor $3$. CP-C-driven THz waveforms appear negligible. These results qualitatively agree with those of the LC model numerically integrated from Eq. (\ref{J}) (solid curves) and providing a factor $\sim 2.3$ in the CP-S field strength, thus a factor $\sim 5$ in its THz power, which is consistent with \cite{Meng:apl:109:131105}. As previously mentioned, discrepancies in the CP configurations are due to an overestimation of the photocurrent efficiency over a full electron density step $\delta N$. Such discrepancies are, however, quite limited for high enough values of $r \geq 0.1$. The insets detail the resulting polarization states, remaining linear in LP-P, but becoming slightly elliptical when they are driven by photocurrents as reported in \cite{Fedorov:ppcf:59:014025}. Departing from both Eqs. (\ref{PCCP}) and (\ref{SpCPS}), this slight ellipticity in the CP-S case results from the pulse envelope and the finite duration of the ionization events (small rotation of the polarization ellipse). For comparison, Figs. \ref{Fig2}(c,d,e) show the evolution in the electron density and THz pulse components for CP-S pulses with different intensity ratios $r = 0.01,0.1,0.5$. We can observe that the agreement between the semi-analytical and numerical computation of the THz pulse components becomes degraded when the value of $r$ is decreased to zero.

\ack
This work was supported by the ANR/ASTRID Project ``ALTESSE'' $\#$ ANR-15-ASTR-0009 and performed using HPC resources from GENCI (Grant $\#$ A0080507594). D.B. and V.V. acknowledge the Research Council of Lithuania for funding this research by the grant No. S-MIP-19-46 and support from the Laserlab-Europe EU-H2020 871124. S.S. acknowledges support by the Qatar National Research Fund through the National Priorities Research Program (Grant $\#$ NPRP 12S-0205-190047). I.B. is thankful for support from the Deutsche Forschungsgemeinschaft (DFG; BA4156/4-2) and from the
German Research Foundation under Germany’s Excellence Cluster PhoenixD (EXC 2122 Project ID 390833453).

\section*{References}

\bibliographystyle{unsrt}
\bibliography{references}
 
\end{document}